\documentclass[prd,a4paper,onecolumn,titlepage,superscriptaddress,showpacs,showkeys,preprintnumbers,nofootinbib]{revtex4-1}
\usepackage[utf8]{inputenc}
\usepackage{amssymb}
\usepackage{dcolumn}
\usepackage{slashed}
\usepackage{mathtools}
\usepackage{multirow}
\usepackage[colorlinks = true, linkcolor = blue, urlcolor  = blue, citecolor = blue,  anchorcolor = green,  hyperindex = true, hyperfigures]{hyperref}

\def\rme{{\mathrm{e}}}
\def\GX#1{G_{\rm{#1}}}
\newcommand{\bra}[1]{\langle #1|}
\newcommand{\ket}[1]{|#1\rangle}
\if@mathematic
\def\vec#1{\ensuremath{\mathchoice
    {\mbox{\boldmath$\displaystyle\mathbf{#1}$}}
    {\mbox{\boldmath$\textstyle\mathbf{#1}$}}
    {\mbox{\boldmath$\scriptstyle\mathbf{#1}$}}
    {\mbox{\boldmath$\scriptscriptstyle\mathbf{#1}$}}}}
\else
\def\vec#1{\ensuremath{\mathchoice
    {\mbox{\boldmath$\displaystyle#1$}}
    {\mbox{\boldmath$\textstyle#1$}}
    {\mbox{\boldmath$\scriptstyle#1$}}
    {\mbox{\boldmath$\scriptscriptstyle#1$}}}}
\fi
\begin{document}


\preprint{CP3-Origins-2017-018, HIM-2017-03, MITP/17-029, TIFR/TH/17-21}

\title{Iso-vector axial form factors of the nucleon in two-flavour lattice QCD} 

\newcommand{\affKPH}{\affiliation{PRISMA Cluster of Excellence and Institut f\"ur Kernphysik, \\ University of Mainz, Johann-Joachim-Becher-Weg~45, 55099~Mainz, Germany}}
\newcommand{\affHIM}{\affiliation{Helmholtz Institute Mainz, University of Mainz, 55099~Mainz, Germany}}

\author{S.~Capitani}\affKPH\affHIM\affiliation{Institut für Theoretische Physik, Goethe-Universität Frankfurt, Max-von-Laue-Str.~1, 60438~Frankfurt~am~Main, Germany}
\author{M.~Della~Morte}\affiliation{CP$^3$-Origins \& Danish IAS, University of Southern Denmark, Campusvej~55, 5230~Odense~M, Denmark}
\author{D.~Djukanovic}\email{d.djukanovic@him.uni-mainz.de}\affKPH\affHIM
\author{G.M.~von~Hippel}\email{hippel@uni-mainz.de}\affKPH
\author{J.~Hua}\affKPH\affHIM
\author{B.~J\"ager}\affiliation{CP$^3$-Origins \& Danish IAS, University of Southern Denmark, Campusvej~55, 5230~Odense~M, Denmark}\affiliation{Department of Physics, College of Science, Swansea University, Swansea~SA2~8PP, United~Kingdom}\affiliation{ETH Zürich, Institute for Theoretical Physics, Wolfgang-Pauli-Str.~27, 8093~Zürich, Switzerland}
\author{P.M.~Junnarkar}\affHIM\affiliation{Tata Institute of Fundamental Research (TIFR), Homi~Bhabha~Rd., 400005~Mumbai, India}
\author{H.B.~Meyer}\email{meyerh@uni-mainz.de}\affKPH\affHIM
\author{T.D.~Rae}\affKPH
\author{H.~Wittig}\email{wittig@uni-mainz.de}\affKPH\affHIM


\begin{abstract}
We present a lattice calculation of the nucleon iso-vector
axial and induced pseudoscalar form factors on the CLS ensembles
using $N_{\rm f}=2$ dynamical flavours of non-perturbatively
$\mathcal{O}(a)$-improved Wilson fermions and an
$\mathcal{O}(a)$-improved axial current together with the pseudoscalar density.
Excited-state effects in the extraction of the form factors are treated using a variety
of methods, with a detailed discussion of their respective merits.
The chiral and continuum extrapolation of the results
is performed both using formulae inspired by Heavy Baryon Chiral Perturbation Theory
(HBChPT) and a global approach to the form factors based on a chiral effective
theory (EFT) including axial vector mesons.
Our results indicate that careful treatment of excited-state effects
is important in order to obtain reliable results for the axial form factors
of the nucleon, and that the main remaining error stems from the systematic
uncertainties of the chiral extrapolation.
As final results, we quote
$g_{\rm A} = 1.278 \pm 0.068\genfrac{}{}{0pt}{1}{+0.000}{-0.087}$,
$\langle r_{\rm A}^2\rangle = 0.360 \pm 0.036\genfrac{}{}{0pt}{1}{+0.080}{-0.088}~\mathrm{fm}^2$, and 
$g_{\rm P} = 7.7 \pm 1.8 \genfrac{}{}{0pt}{1}{+0.8}{-2.0}$ for the axial charge,
axial charge radius and induced pseudoscalar charge, respectively,
where the first error is statistical and the second is systematic.
\end{abstract}


\maketitle


\section{Introduction}

The structure of the nucleon is of fundamental importance in
characterizing matter at subatomic length scales.  Nucleon structure
can be studied experimentally using the electroweak gauge bosons
($\gamma$, $Z$, $W^\pm$) as probes.  In many cases, these interactions
must be understood quantitatively in order to interpret precision
experiments searching for new physics.

The interaction of an electroweak gauge boson with the nucleon is parameterized
by form factors. Specifically, the photon couples via the electromagnetic current,
while the $W^\pm$ boson couples to the left-handed component $\bar q\gamma_\mu(1-\gamma_5)q$
of the quarks with weak-isospin charge factors. While the electromagnetic
form factors are well determined, the matrix elements of the axial current $\bar q\gamma_\mu\gamma_5q$
are less precisely known.
Focusing on the light-quark contribution, 
the nucleon matrix elements of the iso-vector axial current
are encoded in the axial and induced pseudoscalar form factors.
The axial charge of the nucleon, defined as the axial form factor
at zero momentum transfer, can be interpreted as the fractional contribution from
quark and antiquark spins to the nucleon spin and
is known experimentally to an accuracy of two parts per mille through
neutron beta-decay processes \cite{Olive:2016xmw}. 
The momentum-transfer dependence of the axial form factor, which
can be related to the transverse densities of helicity-aligned minus
anti-aligned quarks and antiquarks in the infinite-momentum frame
\cite{Burkardt:2002hr},
is much less well known. A recent analysis~\cite{Meyer:2016oeg} assigns an uncertainty
of about twenty percent to the axial charge radius, which is given by
the slope of the axial form factor at $Q^2=0$ (see Eq.~(\ref{Eq:radius}) below).
The axial form factor is accessible primarily via neutrino scattering
off the nucleon \cite{Kitagaki:1990vs,Bodek:2007vi,Meyer:2016oeg}, and,
at low momentum transfer, 
via the electro-production of charged  pions~\cite{Liesenfeld:1999mv,Fuchs:2003vw}.
There is a tension between the values of the axial radius $\langle r_{\rm A}^2 \rangle$ 
obtained by these two experimental techniques~\cite{Bernard:2001rs}.
The induced pseudoscalar form factor, which is related to the pion-nucleon
form factor through the Goldberger-Treiman relation
\cite{Schindler:2006it,Alexandrou:2007xj},
is measured experimentally in muon-capture processes on the proton
\cite{Wright:1998gi,Winter:2011yp}
and has recently been determined at the seven percent level
\cite{Andreev:2012fj,Andreev:2015evt}.

Lattice QCD determinations of nucleon form factors have a long tradition~\cite{Martinelli:1988rr}.
They are based on evaluating two- and three-point functions in four-dimensional Euclidean space
in the path integral formalism with the help of importance-sampling Monte-Carlo techniques.
Calculations of the nucleon axial charge
\cite{Yamazaki:2008py,Green:2012ud,Capitani:2012gj,Horsley:2013ayv,Bhattacharya:2013ehc,
Bali:2014nma,Chambers:2014qaa,Abdel-Rehim:2015owa,Bhattacharya:2016zcn,Bouchard:2016heu,Yoon:2016jzj,
Liang:2016fgy,Berkowitz:2017gql}
have  tended to yield lower values than the experimental one at phy\-si\-cal 
quark masses. This is widely believed to be due to a failure to
properly account for excited-state contributions in the lattice simulations
\cite{Lin:2012ev,Capitani:2012gj,vonHippel:2016wid},
although lattice cutoff effects, finite-size effects \cite{Yamazaki:2008py,Horsley:2013ayv} and
even finite-temperature effects \cite{Green:2012ud} must be kept under control as well.
Lattice studies of the momentum dependence of the axial form factors are
not as numerous yet, but have become more common recently
\cite{Yamazaki:2009zq,Bratt:2010jn,Alexandrou:2010hf,Meyer:2016kwb,Alexandrou:2017msl,
Green:2017keo,Alexandrou:2017hac,Rajan:2017lxk}.
Since  the axial form factor of the nucleon is an important
 source of uncertainty in determining the neutrino flux in 
long-baseline neutrino experiments \cite{Mosel:2016cwa,Meyer:2016oeg}, an accurate
QCD prediction from the lattice is now particularly timely.

This paper is structured as follows: we describe our general lattice setup
in section \ref{sec:lattice}, and give details on our treatment of the
excited-state contaminations in section \ref{sec:excited}. Our results
for the axial form factor are presented in section \ref{sec:GA}, and
for the induced pseudoscalar form factor in section \ref{sec:GP}. We discuss
different ways of performing the chiral and continuum extrapolation of our
results in section \ref{sec:chiral}, and conclude with a discussion of
our findings and their implications in section \ref{sec:discussion}.

A complete set of our results for the form factors on all lattice ensembles
used is given in \ref{app:tables}.


\section{Lattice set-up}
\label{sec:lattice}

\subsection{Observables and correlators}

We employ a Euclidean notation throughout.
The matrix element of the local iso-vector axial current
$A^{a}_{\mu}(x) = \bar\psi \gamma_\mu\gamma_5\frac{\tau^a}{2}\psi$
between single-nucleon states can be  parameterised by
the axial form factor $\GX{A}(Q^2)$ and induced pseudoscalar
form factor $\GX{P}(Q^2)$ as
\begin{eqnarray}\label{Eq:Amu_mat}
&& \bra{N(p^{\prime},s^\prime)} A^{a}_{\mu}(0) \ket{N(p,s)} = \\ \ \nonumber
&&  \bar{u}(p^{\prime}) \bigg( \gamma_\mu \gamma_5 \GX{A}(Q^2) -i \gamma_5 \frac{Q_\mu}{2M_N} \GX{P}(Q^2)\bigg) 
\frac{\tau^a}{2} u(p),
\end{eqnarray}
where $Q_\mu=( iE_{\vec p'}-iE_{\vec p},\;\vec q)$, $\vec q=\vec p'-\vec p$, $u(p)$ is an isodoublet Dirac spinor with momentum $p$,
$\gamma_\mu$ is a Dirac matrix, and $M_N$ is the nucleon mass.
The square of the four-momentum transferred to the nucleon
via its interaction with the iso-vector axial current is given by
\begin{align}
Q^2 =   (\vec{p}^\prime - \vec{p})^2  -(E_{\vec{p}^\prime} - E_{\vec{p}})^2 .
\end{align}
In this work, the axial and induced pseudoscalar form factors
are computed for space-like momentum transfers $Q^2>0$.
The axial form factor admits a Taylor expansion at low $Q^2$ given by
\begin{equation}\label{eq:GAtaylor}
\GX{A}(Q^2) = g_{\rm A} \bigg(  1 - \frac{1}{6} \langle r^2_{\rm A} \rangle \ Q^2 + \mathcal{O}(Q^4) \bigg),
\end{equation}
where $g_{\rm A} = \GX{A}(Q^2=0)$ is the nucleon axial charge
and $\langle r^2_{\rm A} \rangle$ is the square of the axial charge radius
of the nucleon,
\begin{equation}\label{Eq:radius}
 \langle r^2_{\rm A} \rangle  = -\frac{6}{g_{\rm A}} \left.\frac{\partial \GX{A}(Q^2)}{\partial Q^2}\right|_{Q^2=0}.
\end{equation}
The pseudoscalar coupling is defined by
\begin{equation}
g_{\rm P}\equiv \frac{m_\mu}{2m_N} \GX{P}(Q_*^2),\label{Eq:smallgp}
\end{equation}
with $Q_*^2=0.88m_\mu^2$ the momentum transfer relevant to muon capture
with the nucleon at rest \cite{Measday:2001yr}.
The nucleon matrix element of the iso-vector axial current
is related to that of the pseudoscalar current via the
chiral Ward identity in two-flavour QCD also known as
the partially conserved axial current (PCAC) relation,
\begin{equation}
\label{Eq:pcac}
\partial_\mu A^{a}_{\mu}(x) = 2\, m_{q} P^a(x),
\end{equation}
where $P^a(x)=\bar\psi \gamma_5 \frac{\tau^a}{2}\psi$ is the pseudoscalar density and $m_q$ is the
average quark mass in the isospin limit. 
The matrix element of the pseudoscalar density between
single-nucleon states is given by
\begin{equation}
\label{Eq:Fpdef}
 m_q \bra{N(p^{\prime},s^\prime)} P^a(0) \ket{N(p,s)} =  m_q F_{\rm P}(Q^2)\big( \bar{u}(p^\prime) \gamma_5 \frac{\tau^a}{2} u(p) \big),
\end{equation}
where $F_{\rm P}(Q^2)$ is the pseudoscalar form factor.
It is related to the pion-nucleon form factor
$G_{\pi N}(Q^2)$ through the relation~\cite{Schindler:2006it}
\begin{equation}\label{Eq:FP_GpiN}
m_q F_{\rm P}(Q^2) =  \frac{m^2_\pi F_\pi}{m^2_\pi + Q^2} G_{\pi N} (Q^2),
\end{equation}
 $F_\pi=92.4{\rm\,MeV}$ being the pion decay constant.
Taking the matrix element of the PCAC relation in Eq.~(\ref{Eq:pcac})
between single-nucleon states provides another relation between
the form factors in Eqs.~(\ref{Eq:Amu_mat}) and (\ref{Eq:Fpdef}),
\begin{equation} \label{Eq:pcac_mat}
2 M_N \GX{A}(Q^2) - \frac{Q^2}{2M_N}\GX{P}(Q^2) = 2 m_q F_{\rm P}(Q^2).
\end{equation}
In this work, we use this relation to study the form factors
in Eq.~(\ref{Eq:Amu_mat}).
The induced pseudoscalar form factor has a pole at the pion mass,
as dictated by chiral symmetry breaking via the Goldberger-Treiman
\cite{Goldberger:1958tr,Goldberger:1958vp,Nambu:1960xd}
relation $G_{\pi N}(Q^2)F_\pi=\GX{A}(Q^2)M_N$ for $Q^2\to 0$.

\subsection{Simulation details}

The eleven ensembles used in this work are identical to those used in our
calculation of electromagnetic form factors \cite{Capitani:2015sba},
and the reader is referred to Table~I of Ref.~\cite{Capitani:2015sba}
for details\footnote{Note that the number of measurements is identical to Ref.
\cite{Capitani:2015sba}.}. There are three lattice spacings,
$a=0.079,\;0.063$ and $0.050{\rm\,fm}$,
the lightest pion mass is 190\,MeV and the physical volumes satisfy $m_\pi L\geq 4.0$.
The ensembles, which were generated as part of the CLS (Coordinated
Lattice Simulations) initiative, employ $N_{\rm f}=2$ flavours of
non-perturbatively $\mathcal{O}(a)$-improved Wilson fermions.
The Monte Carlo simulations were performed using the deflation-accelerated
DD-HMC
\cite{Luscher:2005rx,Luscher:2007es}
and MP-HMC
\cite{Marinkovic:2010eg}
algorithms.
The value of the improvement coefficient $c_{\rm sw}$ was determined
non-perturbatively in Ref.~\cite{Jansen:1998mx}.

The setup for our lattice determination of the nucleon matrix element
of the iso-vector axial current and pseudoscalar density is likewise
very similar to the one we used in the case of the electromagnetic
current \cite{Capitani:2015sba}. 
We will always be evaluating the third isospin component of the
axial current and pseudoscalar density on the proton, and therefore drop isospin indices from now on.
The nucleon two-point function is computed as
\begin{equation}
\label{Eq:C2pt}
C_2(\vec{p},t) = \sum_{\vec{x}} \rme^{i\vec{p}\cdot\vec{x}}\, \Gamma_{\beta\alpha}\,\langle \Psi^\alpha(\vec{x},t)\overline{\Psi}^\beta(0)\rangle, 
\end{equation}
where $\Psi^\alpha(\vec{x},t)$ denotes the nucleon interpolating
operator constructed as
\begin{equation}
\Psi^\alpha(x) =  \epsilon_{abc} \big( \tilde{u}^T_a(x) C \gamma_5 \tilde{d}_b(x) \big) \tilde{u}^\alpha_c(x)
\end{equation}
using Gaussian-smeared quark fields
\cite{Gusken:1989ad}
\begin{equation}
\label{eq:gauss}
  \widetilde\psi = \left(1+\kappa_{\rm G}\Delta\right)^N \psi\,.
\end{equation}
In Eq.~(\ref{eq:gauss}), the gauge links entering the covariant
three-dimensional Laplacian $\Delta$ have been spatially APE-smeared
\cite{Albanese:1987ds}
in order to reduce the gauge noise and to further enhance the
projection properties onto the nucleon ground state.
Our parameter choices for $\kappa_{\rm G}$ and $N$ correspond to a
smearing radius
\cite{vonHippel:2013yfa}
of around $r_{\rm{sm}}\approx 0.5$\,fm.

The nucleon three-point function is computed with the kinematics
chosen such that the nucleon at the sink is always at rest, i.e. $\vec{p}'=0$.
This ``fixed-sink'' method allows for arbitrary insertion times for the
current operator. 
In this work we consider the three-point functions with the local operator
$\mathcal{O}(\vec{y},t) \in \{ A^I_\mu,  P \}$, schematically represented as
\begin{equation}
\label{Eq:C3pt}
C_{3,\mathcal{O}}(\vec{q},t,t_s) = \sum_{\vec{x},\vec{y}}
\rme^{i\vec{q}\cdot\vec{y}}\Gamma_{\beta\alpha}\langle
\Psi^\alpha(\vec{x},t_s) \mathcal{O}(\vec{y},t)\overline{\Psi}^\beta(0)\rangle\,,
\end{equation}
where  $t_s$ denotes the nucleon source-sink separation,
and $t$ denotes the timeslice of the local operator insertion.

To ensure that all of our observables are $\mathcal{O}(a)$-improved, we use the
renormalised iso-vector axial current including $\mathcal{O}(a)$ improvement,
\begin{equation}
\label{Eq:AmuR}
A^I_\mu(x) = Z_{\rm A} (1 + b_{\rm A} am_q) \left(A_\mu(x) + ac_{\rm A} \partial_\mu P(x) \right)
\end{equation}
where $A_\mu$ and $P$ are the bare local axial current and pseudoscalar
density, respectively, and $m_q$ is the bare subtracted quark mass.
The renormalisation factor $Z_{\rm A}$ and the  improvement coefficient $c_{\rm A}$
have been determined non-perturbatively in Refs.~\cite{DellaMorte:2008xb} and
\cite{DellaMorte:2005aqe}, respectively, and the mass-dependent improvement coefficient
$b_{\rm A}$ was computed in tadpole-improved perturbation theory
in Ref.~\cite{Sint:1997jx}. The pseudoscalar density is automatically
$\mathcal{O}(a)$ improved.

The projection matrix $\Gamma$ is chosen as
\begin{equation}
\Gamma = \frac{1}{2}(1 + \gamma_0)(1 + i \gamma_5 \gamma_3) \label{Eq:proj_matrix}
\end{equation}
and is identical to the one used in Ref.~\cite{Capitani:2015sba}.
Both three-point and two-point functions are constructed using identical
smearing at source and sink in order to ensure a positive spectral
representation.

The matrix elements of the local operator $\mathcal{O}(\vec{y},t)$
are encoded in the three-point function and can be isolated by constructing
appropriate ratios of the three-point and two-point functions, in which the
normalisation of the interpolating operators cancels.
We use the ratio
\begin{equation}
\label{Eq:ratio}
R_{\mathcal{O}}(\vec{q},t,t_s) \equiv \frac{C_{3,\mathcal{O}}(\vec{q},t,t_s)}{C_2(0,t_s)} \sqrt{ \frac{C_2(\vec{q},t_s-t)C_2(\vec{0},t)C_2(\vec{0},t_s)}{C_2(\vec{0},t_s-t)C_2(\vec{q},t)C_2(\vec{q},t_s)}}\,,
\end{equation}
which was found to be particularly effective in isolating the
ground-state matrix elements \cite{Alexandrou:2008rp}
in the asymptotic limit $t,t_s\to\infty$,
where the single-nucleon state dominates.


\section{Analysis of excited state contamination}
\label{sec:excited}

For the iso-vector axial current $A^I_\mu(x)$ of Eq.~(\ref{Eq:AmuR}) and using the projection matrix of Eq.~(\ref{Eq:proj_matrix}),
the asymptotic values $R^0_{A_\mu}(\vec{q})$ of the ratios can be shown
to have the following form:
\begin{align}\label{Eq:GAGP}
R_{A_{0}}(\vec{q},t,t_s)  &\xrightarrow[t,t_s \rightarrow \infty]{} R^0_{A_0}(\vec{q})\\\nonumber &=  \frac{q_3}{\sqrt{2 E_q(M_N+E_q)}} \bigg( \GX{A}(Q^2) + \frac{M_N - E_q}{2M_N} \GX{P}(Q^2)\bigg)\,, \\ \nonumber \vspace{0.2cm}
R_{A_{k}}(\vec{q},t,t_s)  &\xrightarrow[t,t_s \rightarrow \infty]{} R^0_{A_k}(\vec{q}) \\\nonumber &= \frac{i}{\sqrt{2 E_q(M_N+E_q)}} \bigg( \big(M_N+E_q \big)  \GX{A}(Q^2) \delta_{3k}  - \frac{\GX{P}(Q^2)}{2M_N} q_3 q_k \bigg)\,, 
\end{align}
where $E_q$ is the energy of a nucleon with momentum $\vec{q}$
as given by the lattice dispersion relation.

The ratio of the pseudoscalar density $R^0_P(\vec{q})$ also provides access
to the axial and induced pseudoscalar form factors via the PCAC relation in
Eqs.~(\ref{Eq:pcac}) and (\ref{Eq:pcac_mat}), with an asymptotic value given by
\begin{equation}
2 m_q R_P(\vec{q},t,t_s) \xrightarrow[t,t_s \rightarrow \infty]{} 2 m_q R^0_{P}(\vec{q}) =2 M_N R^0_{A_0}(\vec{q})
\end{equation}
We note that the PCAC relation implies that the product of the bare
quark mass and the pseudoscalar density is renormalised by the
renormalisation constant $Z_{\rm A}$ of the axial current.
However, in the course of our analysis we found that the temporal component
$A_0$ of the axial current was too noisy and too affected by excited-state contributions
to be included in the determination of the form factors.

In the asymptotic ratios $R^0_{\rm A,P}$ of the axial current
and pseudoscalar density, the axial and induced pseudoscalar form factors
$\GX{A,P}$ appear in linear combinations, from which they can be determined
by solving the (generally overdetermined) linear system in Eq.~(\ref{Eq:GAGP}).
For a given four-momentum transfer $Q^2$, this is done by minimizing the
least-squares function
\begin{equation}
\label{Eq:chisq}
\chi^2 = \sum^N_{i,j} \big( \mathbf{R} - \mathbf{M} \ \mathbf{G} \big)_i \  \left(\sigma^{-2}\right)_{ij} \ \big(\mathbf{R} - \mathbf{M} \ \mathbf{G} \big)_j \,,
\end{equation}
where $\sigma^2$ is the covariance matrix of the ratios $R_i$ and
\[ 
\mathbf{R} =\begin{pmatrix} R_1 \\    \vdots\\   R_N\\  \end{pmatrix}, \quad 
\mathbf{M} =\begin{pmatrix} M_{1,\rm A} & M_{1,\rm P} \\    \vdots & \vdots\\   M_{N,\rm A} & M_{N,\rm P}\\  \end{pmatrix}, \quad
 \mathbf{G} = \begin{pmatrix}  \GX{A} \\ \\ \GX{P}\end{pmatrix}.
\]
At each four-momentum transfer $Q^2$, the ratios for those individual
three-momentum vectors $\vec{q}$ which are related by an exact symmetry
of the lattice are averaged, and the resulting averaged ratios are
combined into the vector $\mathbf{R}^\top = (R_1 \ldots R_N)$.
In Table \ref{tab:comps}, 
we list, for each momentum transfer, the number $N$ of ratios coming from the
various components of the axial current and the pseudoscalar density
which remain after averaging over equivalent momenta.
The kinematic factors associated with each of the averaged ratios are
represented by the rectangular matrix $\mathbf{M}$ of size $(N \times 2)$.

\begin{table}[t]
\caption{Numbers of momentum-averaged components of the
         axial current $A_{\mu}(x)$ and pseudoscalar density $P(x)$ available
         for solving the linear system in Eq.~(\ref{Eq:chisq}) at various
         momentum transfers.\vspace{16pt}}
{
\renewcommand{\arraystretch}{2.0}
\begin{tabular}{ccccc}\hline\hline
        $\frac{\vec{q}^2L^2}{4 \pi^2}$ & $A_{1,2}$ & $A_3$ & $P$& N \\
        \hline
        0 & - & 1 & - & 1  \\ 
        \hline
        1 & 0 & 2 & 1 & 3 \\        
        \hline
        2 & 1 & 2 & 1 & 4 \\        
        \hline
        3 & 1 & 1 & 1 & 3 \\        
        \hline
        4 & 0 & 2 & 1 & 3 \\
         \hline
        5 & 1 & 3 & 2 & 6 \\
         \hline
        6 & 2 & 2 & 2 & 6 \\\hline\hline
\end{tabular}
\label{tab:comps}
}
\end{table}

In obtaining the form factors from the measured ratios,
we can proceed in two different ways, which differ
by the order in which the extraction of the asymptotic behaviour and
the reduction into form factors are performed:
\begin{enumerate}
\item \textit{Computing effective form factors:}
In this approach, the linear system resulting from Eq.~(\ref{Eq:chisq}) is solved
for each operator insertion time $t$, source-sink separation $t_s$, and
four-momentum transfer $Q^2$, yielding the so-called effective form factors
$G^{\text{eff}}_{\rm A,P}(Q^2,t,t_s)$.
The effective form factors still contain short-distance contributions
from multi-particle and excited states, which need to be accounted for
in order to determine the ground-state form factors; we will discuss
the methods used for this purpose below. 
This method allows for the visualisation of the
approach of $G^{\text{eff}}_{\rm A,P}(Q^2,t,t_s)$ towards the ground-state
form factors $G_{\rm A,P}(Q^2)$ as $t,t_s\to\infty$
(cf. Fig.~\ref{Fig:effGA}).
\item \textit{Computing asymptotic ratios:}
In this approach, the excited-state analysis is first applied to the
vector of averaged ratios $\mathbf{R}(Q^2,t,t_s)$ in order to obtain
asymptotic ratios $\mathbf{R}^0(Q^2)$ for $t,t_s\to\infty$.
The linear system resulting from (\ref{Eq:chisq}) is then solved on these asymptotic
ratios, which then directly yields the ground-state form factors
$G_{\rm A,P}(Q^2)$.
\end{enumerate}

The determination of the asymptotic quantities from their effective
counterparts is rendered non-trivial by the combination of the exponentially
decaying signal-to-noise ratio of baryonic correlation functions at large time
separations and the presence at short time separations of contributions from
excited and multi-particle states.
These excited-state contributions vanish exponentially and give rise
to corrections of the form
\begin{align}\label{Eq:esc}
G^{\text{eff}}_{\rm A,P}(Q^2,t,t_s) &= G_{\rm A,P}(Q^2) \\ &\times \bigg( 1 + \mathcal{O}(e^{-\Delta t}) + \mathcal{O}(e^{-\Delta^{\prime} (t_s-t)}) \bigg)\nonumber,
\end{align}
where $\Delta$ and $\Delta^{\prime}$ are the energy gaps between the ground
and excited states of the initial and final-state nucleons.
A corresponding relation holds between the ratios $\vec{R}(Q^2,t,t_s)$ and
their asymptotic values $\vec{R}^0(Q^2)$.
While the contributions from excited states can in principle be made
exponentially small by taking both $t$ and $t_s-t$ to be large, the exponential
decrease of the signal-to-noise ratio makes this approach impracticable,
as very high statistics would be required to go significantly beyond
$t_s\sim 1.2$~fm.%
\footnote{However, the use of techniques such as all-mode-averaging
\cite{Blum:2012uh,Shintani:2014vja}
may provide a means to study source-sink separations as large as
$t_s\sim 1.6$~fm with reasonable statistical accuracy
\cite{vonHippel:2016wid}.
}
As previously observed \cite{Capitani:2015sba}, a source-sink separation of
at least $t_s \gtrsim 0.5$~fm is required to achieve ground-state saturation
in the two-point function for single nucleon states with zero momentum.
For nucleon states with non-zero momenta the limitation is even
more severe, and in the case of three-point functions, both $t$ and $t_s-t$
must be made sufficiently large, so that source-sink separations larger than
$t_s > 1.5$~fm would be required to achieve ground-state saturation.
At the currently achievable source-sink separations of $t_s \sim 1 -1.2 $~fm
used in this work, we can therefore not rely on ground-state saturation,
and a systematic analysis of the excited state contributions is necessary.
In our previous work
\cite{Capitani:2012gj,Jager:2013kha,Junnarkar:2014jxa,Capitani:2015sba},
we have found two methods to be particularly useful in studying the
excited-state contributions, namely
\begin{enumerate}
\item[A.] \textit{Summation method:} This method starts from constructing summed
ratios \cite{Maiani:1987by,Bulava:2011yz,Brandt:2011sj,Green:2012ud} at each
four-momentum transfer $Q^2$ and source-sink separation $t_s$. The summed
ratios can be shown to be asymptotically linear in the source-sink separation
$t_s$, with the form factors $\GX{A,P}$ appearing as the slope,
\begin{align}
 S(t_s) &\equiv  \sum^{t_s-1}_{t=1} G^{\rm eff}_{\rm A,P}(Q^2,t,t_s)  
\label{eq:summation} \\
\nonumber &\rightarrow  K(Q^2) + t_s \ G_{\rm A,P}(Q^2) + \ldots,
\end{align}
where $K(Q^2)$ denotes a constant intercept, and the ellipses indicate
neglected subleading contributions of $\mathcal{O}(t_s e^{-\Delta t_s})$
and $\mathcal{O}(t_s e^{-\Delta^{\prime} t_s})$. 
\item[B.] \textit{Two-state fits:} In this method, the excited-state
contributions are explicitly modelled using the ansatz
\begin{align}\label{eq:GenAnsatz}
G^{\rm eff}_{\rm A,P}(Q^2,t,t_s) = G_{\rm A,P}(Q^2) 
    &+ c_1(Q^2)  \ e^{-\Delta t}  \\ \nonumber
    &+ c_2(Q^2) \ e^{-\Delta^\prime (t_s - t )},
\end{align}
where the ground-state form factors $G_{\rm A,P}(Q^2)$ and amplitudes
$c_1(Q^2)$, $c_2(Q^2)$ are determined by fitting Eq.~(\ref{eq:GenAnsatz})
to the data for all source-sink separations $t_s$ and insertion times $t$
at each value of the four-momentum transfer $Q^2$.
In the case of the axial charge, we are also able to determine the
amplitude of the transition from the excited state to the 
excited state, due to the symmetry of the three-point function under
the transformation $t\to(t_s-t)$ -- see Eq.\ (\ref{Eq:gAansatz})
below. We have used the ansatz (\ref{eq:GenAnsatz}) to perform
two-state fits in our previous study of nucleon electromagnetic form
factors~\cite{Capitani:2015sba}. We note that in
Refs.~\cite{Alexandrou:2017hac,Rajan:2017lxk} the term ``two-state
fit'' denotes an ansatz that also includes the excited-to-excited
contribution. In principle, the gaps $\Delta$, $\Delta^\prime$ can be
determined from the fits; in practice, however, we have found the
resulting fits to be unstable, and in order to obtain meaningful
uncertainties in the fit parameters, an explicit ansatz is made for
the gaps. 
On our lattice ensembles, we expect the low-lying energy levels to be separated
typically by several hundred MeV. With our source-sink separations $t_s\gtrsim
1\,$fm, the higher excited states should then be suppressed in the three-point
correlation function. The simplest model for the excited nucleon spectrum
consists of a set of non-interacting multi-hadron states.
In our setup, the initial-state nucleon is moving, which motivates the
ansatz of an $N\pi$ state with the pion at rest for the dominant excited-state contribution,
corresponding to a gap $\Delta = m_\pi$. The final-state nucleon, on the
other hand, is at rest, motivating the ansatz of an $S$-wave $N\pi\pi$ state
with gap $\Delta^\prime=2m_\pi$ for the dominant excited-state contribution.
In Ref.~\cite{Hansen:2016qoz}, the $N\pi$ excited spectrum was investigated 
thoroughly at physical quark masses, including the effects of interactions via the experimentally
known $P$-wave scattering phase. The effect of the interaction on the energy level 
is small, and at the volumes of $m_\pi L\approx 4$ investigated
here, the first excited $N\pi$ state is practically degenerate with 
the $S$-wave $N\pi\pi$ state when interactions are neglected.

\end{enumerate}
Setting the finite-volume energy gaps to the values corresponding to no interactions
between pions and the nucleon may introduce a systematic bias in the two-state fit method.
The summation method, on the other hand, makes no specific assumptions about the values of 
the energy gaps; it only assumes that terms of order  $e^{-\Delta^{(\prime)} t_s}$
can be neglected. 

The summation method thus involves weaker assumptions about the
excited-state contamination than our implementation of the two-state
fit method. On the other hand, both methods neglect terms of order
$e^{-\Delta^{(\prime)} t_s}$ in the spectral representation.
Therefore, in order to assess the stability of the
physics results under variations of the analysis procedure, we apply
both methods in our study of the axial and induced pseudoscalar form
factors $\GX{A,P}(Q^2)$ of the nucleon.


\begin{figure*}
	\centering
	\includegraphics[width=0.48\textwidth]{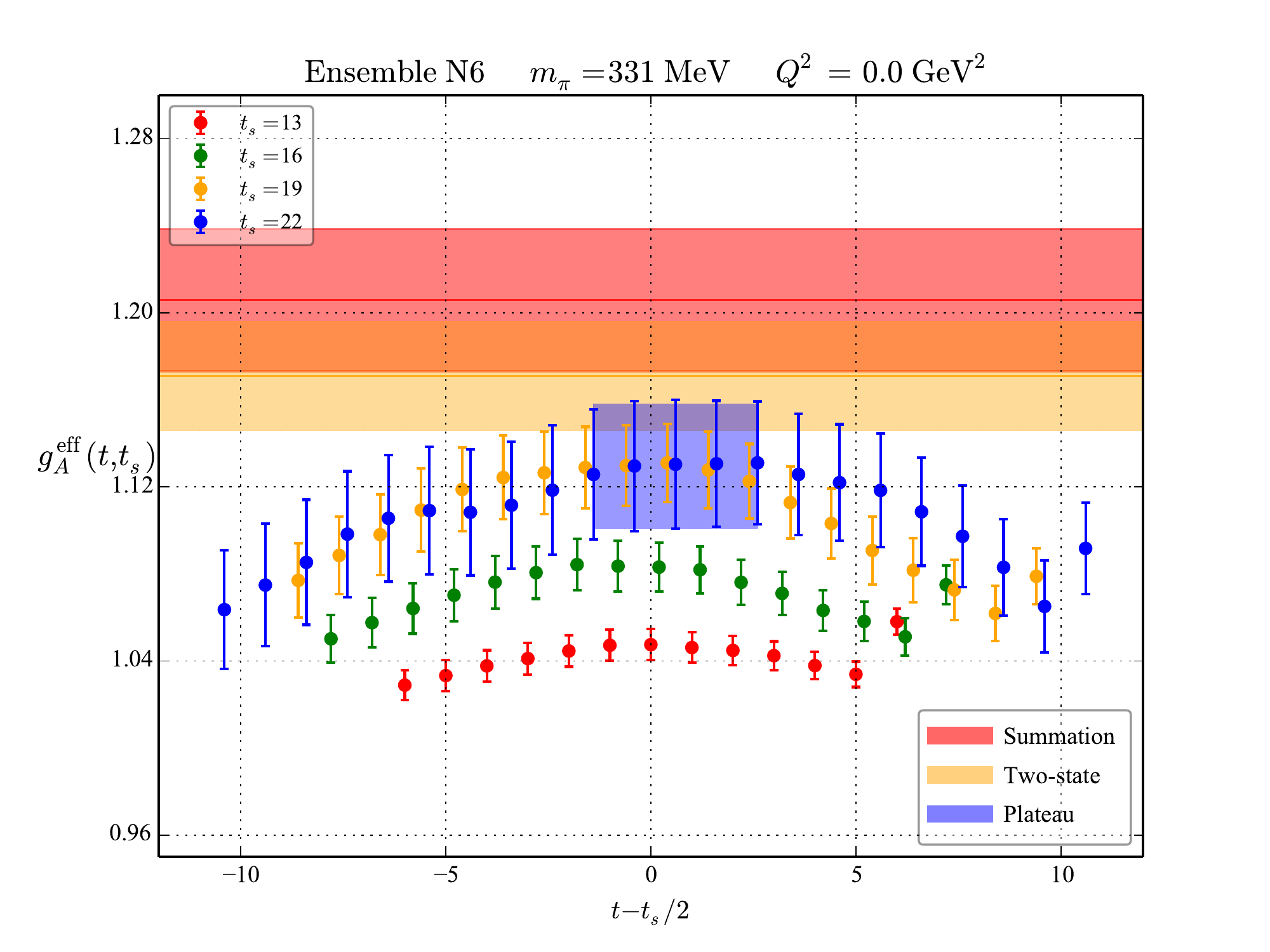}
	\includegraphics[width=0.48\textwidth]{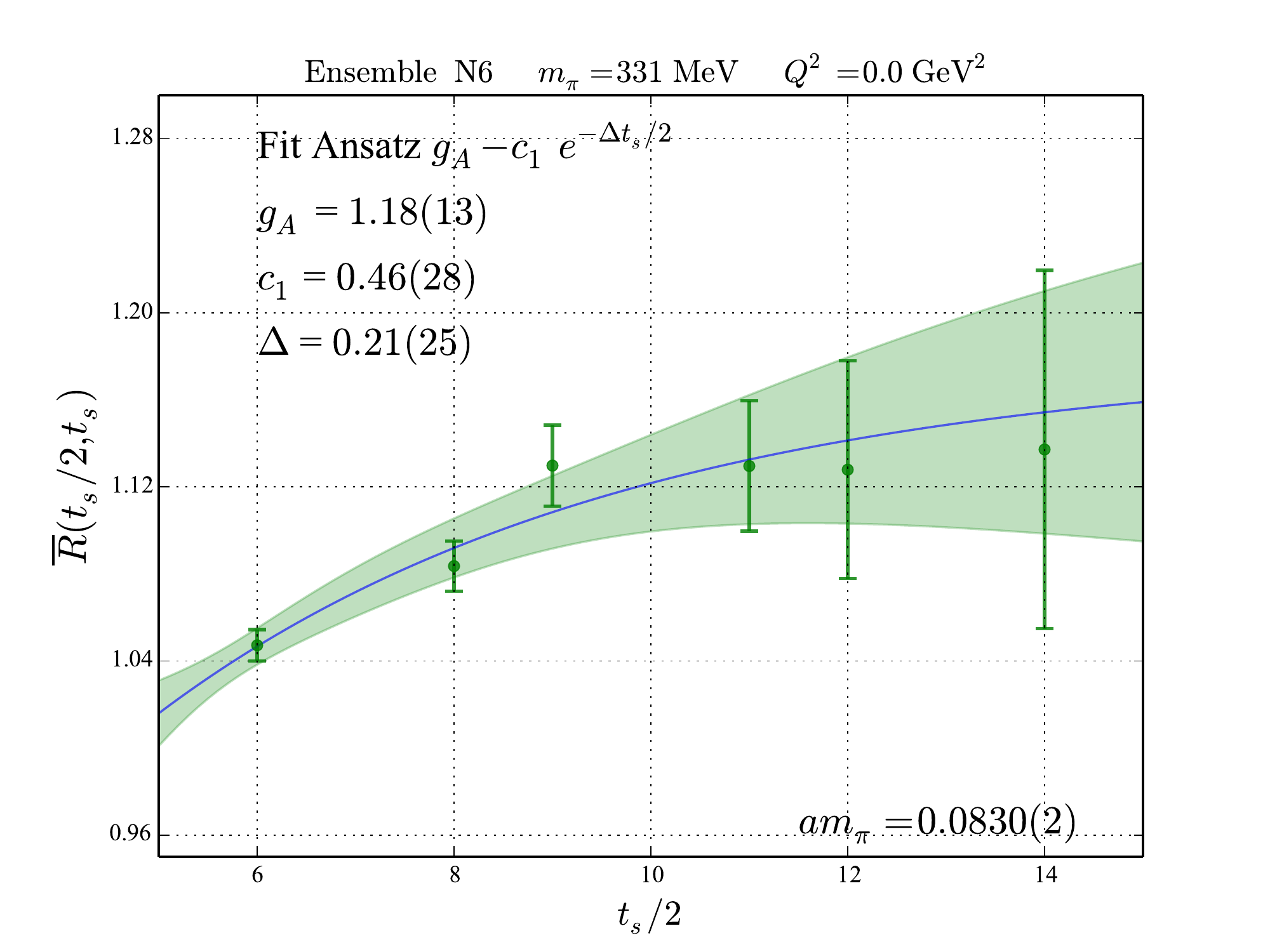}
	\includegraphics[width=0.48\textwidth]{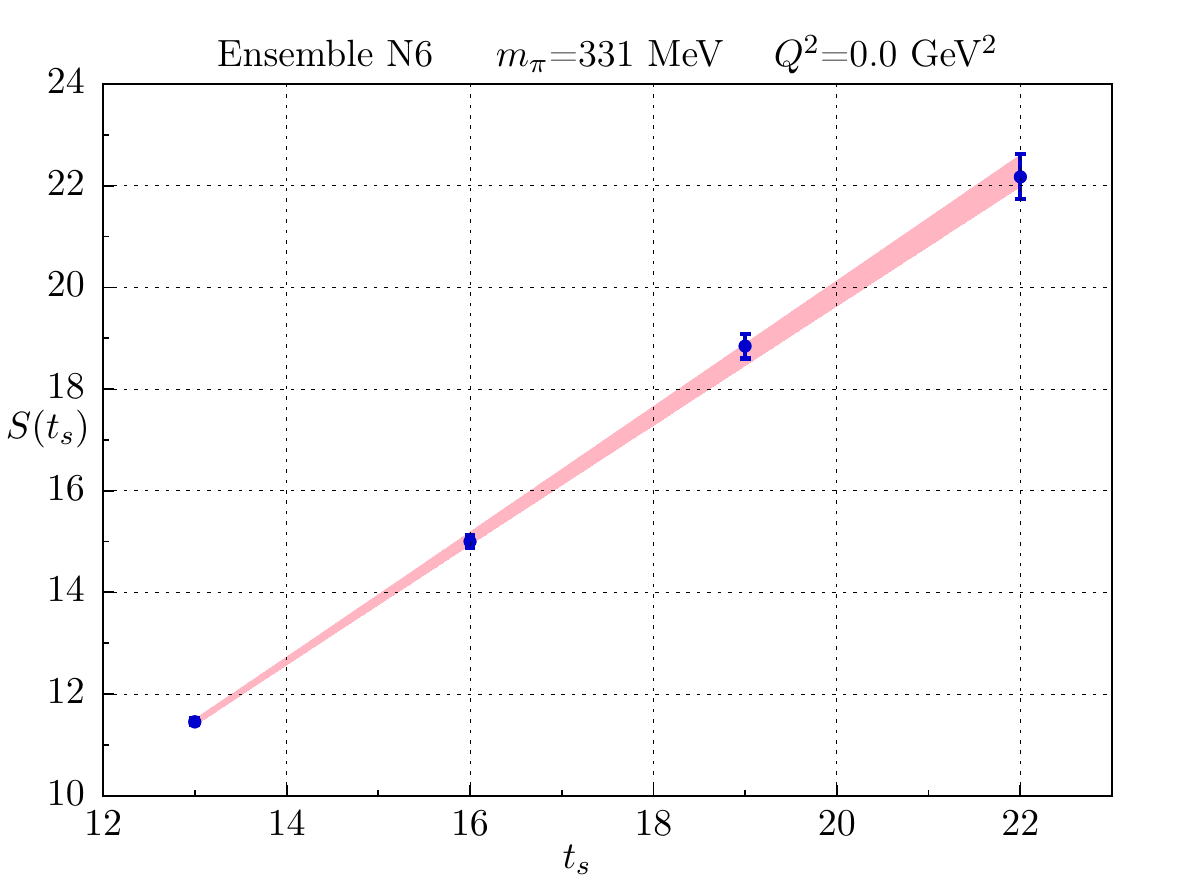}
	\caption{\label{fig:gAN6} Left panel: Effective axial charge
                 $g_{\rm A}^{\rm eff}(t,t_s)$ on Ensemble N6; different
                 source-sink separations $t_s$ are displayed in different colours;
                 also shown are a plateau fit (blue band) at the largest
                 source-sink separation $t_s=22a\approx 1.1$~fm, and the
                 results for $g_{\rm A}$ obtained using the summation
                 method (red band) and a two-state fit (yellow band).
                 Right Panel: Results of plateau fits at different
                 source-sink separations $t_s$ with a fit to the expected
                 $t_s$-dependence. Bottom panel: illustration of the summation method, where 
                 $S(\tau=t_s)$ is the summed insertion (\ref{eq:summation}). 
                 The slope corresponds to the axial charge.}
\end{figure*}

\begin{figure*}
        \centering
        \includegraphics[width=0.48\textwidth]{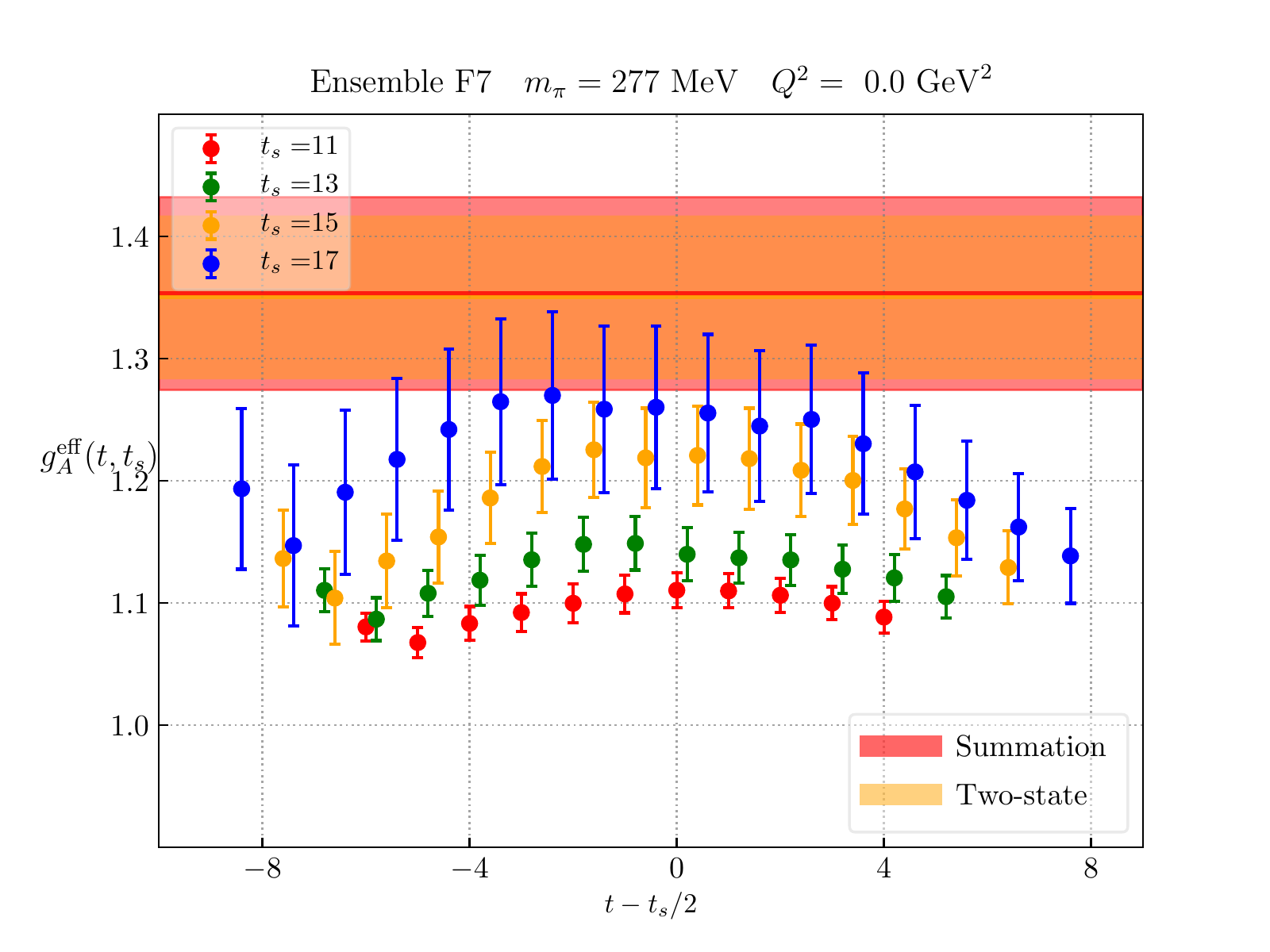}
        \includegraphics[width=0.48\textwidth]{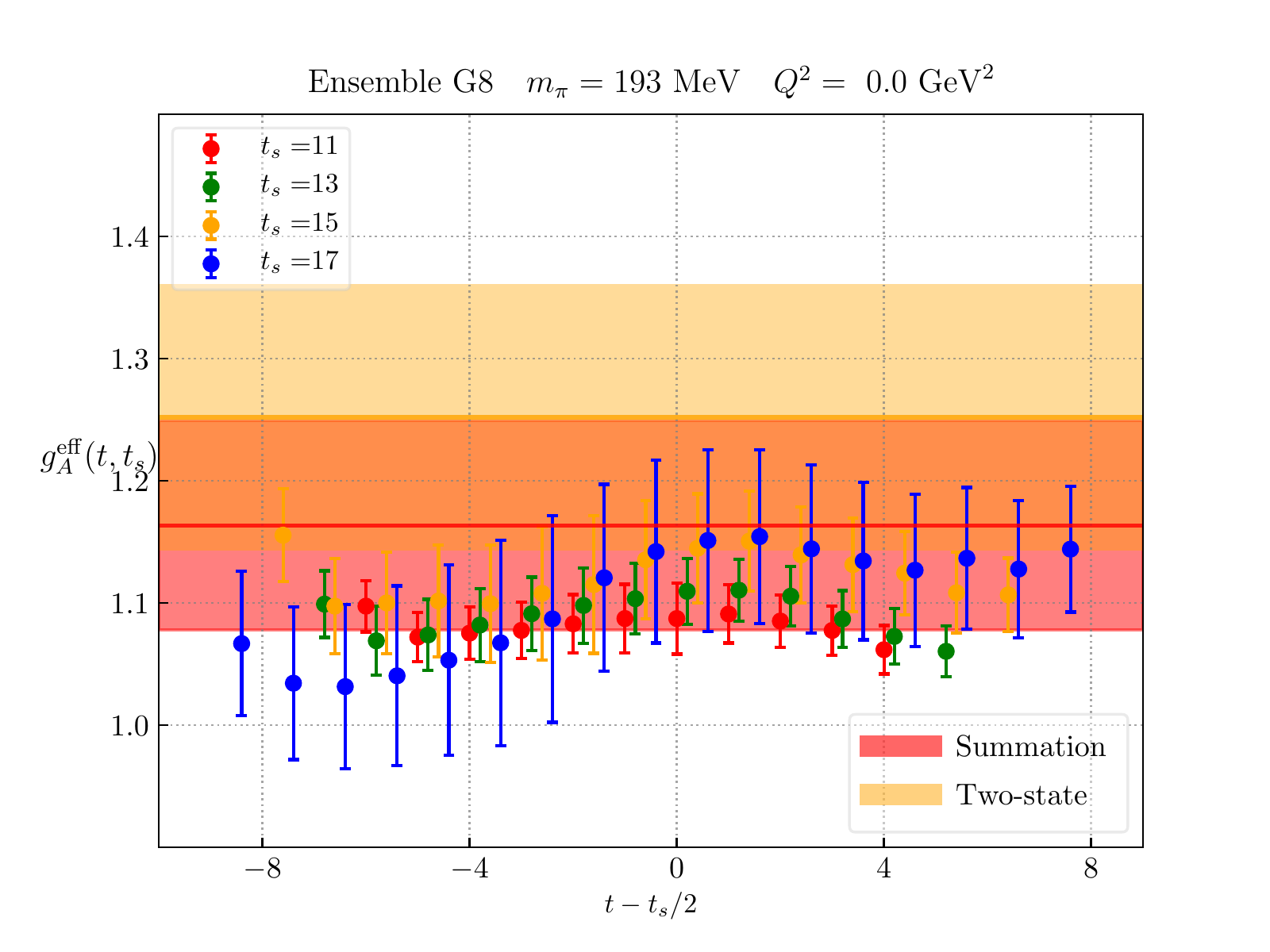}
        \caption{\label{fig:gAG8F7}
        The effective axial charge on the ensembles F7 ($m_\pi=277$\,MeV, left panel)
        and G8 ($m_\pi=193\,$MeV, right panel).
        Different source-sink separations $t_s$ are displayed in different colours;
        the bands represent the result for the axial charge obtained using the
        summation method (red band) and a two-state fit (yellow band)
        to extract ground-state matrix elements.}
\end{figure*}

\begin{figure*}
	\centering
	\includegraphics[width=0.48\textwidth]{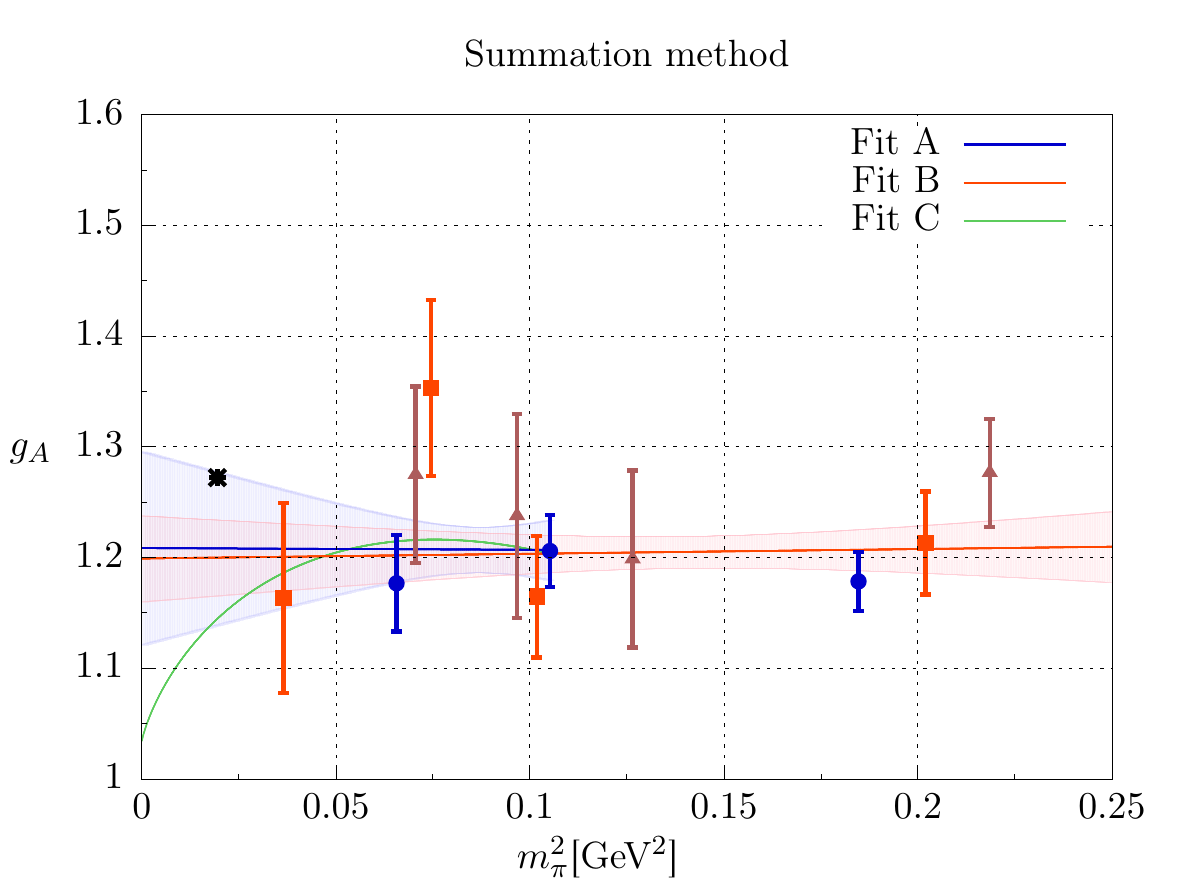}
	\includegraphics[width=0.48\textwidth]{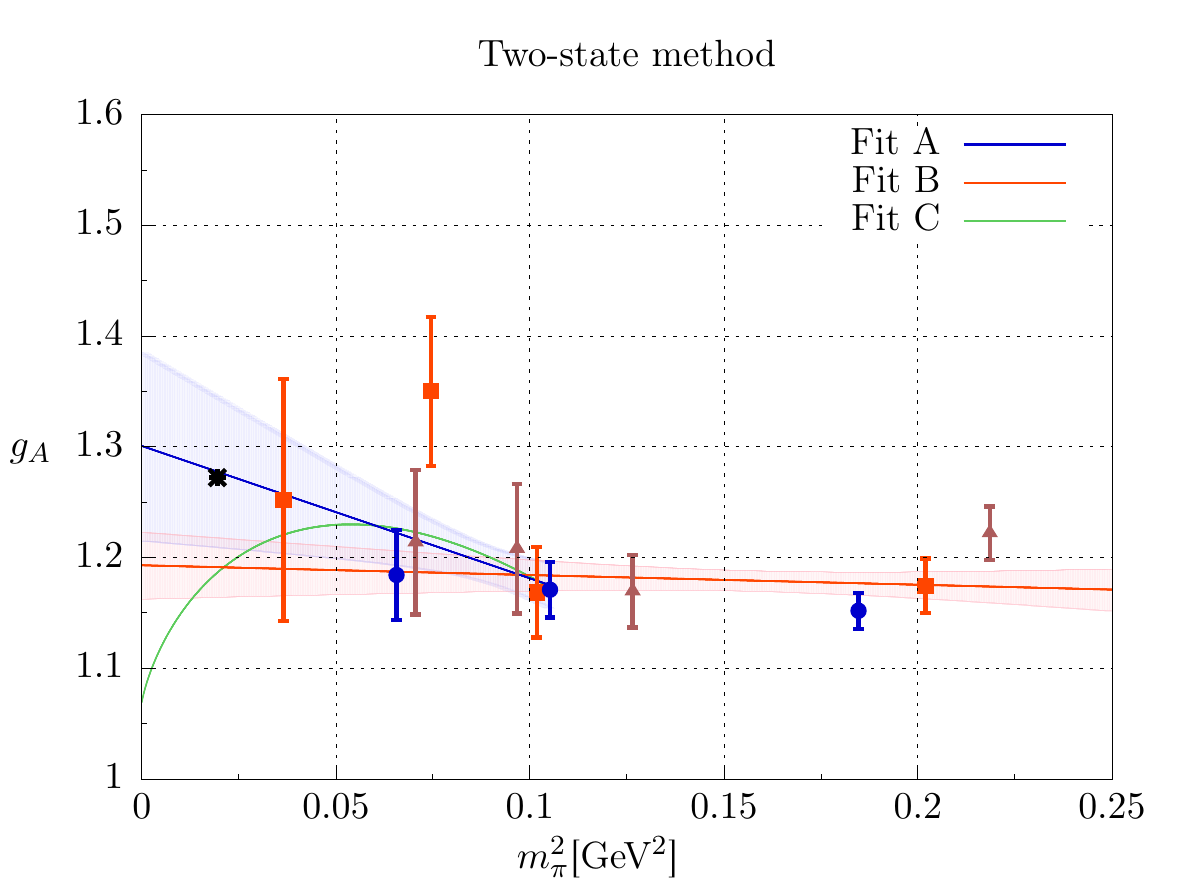}
	\caption{\label{fig:gAextrapolation} Chiral extrapolation of the
                 axial charge $g_{\rm A}$ obtained using the summation
                 method (left panel) and a two-state fit (right panel)
                 to the physical pion mass. 
 Triangles, squares and circles correspond to increasingly fine lattice spacings,
 and the black point represents the phenomenological value of $g_{\rm A}$.
 Fit A is a linear fit with a pion-mass cut $m_\pi\leq 335{\rm\,MeV}$, 
 Fit B is a linear fit with no pion-mass cut, and
 Fit C is based on chiral effective theory via the ansatz of Eq.\ (\ref{eq:xptgafit}).}
\end{figure*}

\section{Isovector axial form factor}\label{sec:GA}

\subsection{Axial charge} 

The axial charge $g_{\rm A}=\GX{A}(0)$ can be determined directly from the
matrix element of the $z$-component of the axial current $A_3$  at zero
momentum transfer where the ratio in Eq.~(\ref{Eq:ratio}) takes the simplified
form
\begin{equation}
g^{\rm eff}_{\rm A}(t,t_s) \equiv R_{A_3}(0,t,t_s) = \frac{C_{A_3}(0,t,t_s)}{C_2(0,t_s)}.
\end{equation}
Since the initial and the final states are identical,
the excited state contributions will be the same at source and sink,
and we expect the effective axial charge $g^{\rm eff}_{\rm A}(t,t_s)$
to approach its asymptotic value in a symmetric fashion.
Moreover, since the nucleon is at rest in the inital and final state,
we expect the dominant excited-state contributions can arise from
the S-wave $N \pi \pi$ multiparticle state, i.e. a nucleon and two
pions at rest, leading to the ansatz $\Delta = \Delta^{\prime} = 2 m_\pi$
for the mass gap. For analytic studies of the excited-state contamination
based on chiral effective theory, see~\cite{Tiburzi:2015tta,Bar:2015zwa,Bar:2016uoj},
and~\cite{Hansen:2016qoz} for a study based on L\"uscher's finite-volume formalism.
In the two-state fits for the axial charge, we therefore use the
fit form
\begin{equation}
g^{\rm eff}_{\rm A}(t,t_s) = g_{\rm A} + \tilde{c}_1 \left(e^{- 2 m_\pi t} + e^{- 2 m_\pi(t_s-t)} \right) + \tilde{c}_2 e^{-2 m_\pi t_s} \,,
\label{Eq:gAansatz}
\end{equation}
where $m_\pi$ is fixed to the pion mass on the ensemble.

\begin{figure*}
        \centering
        \includegraphics[width=0.48\textwidth]{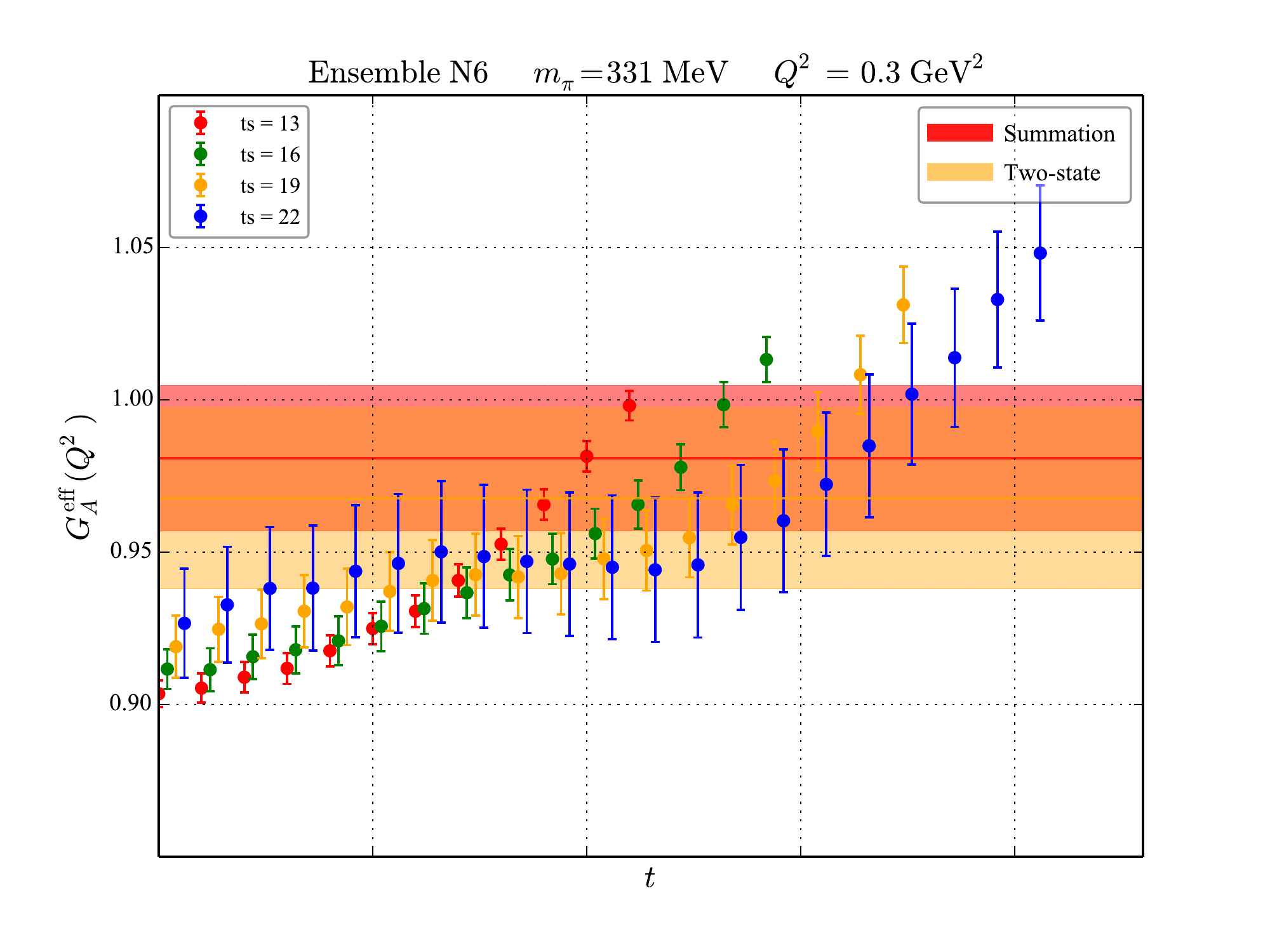}
        \includegraphics[width=0.48\textwidth]{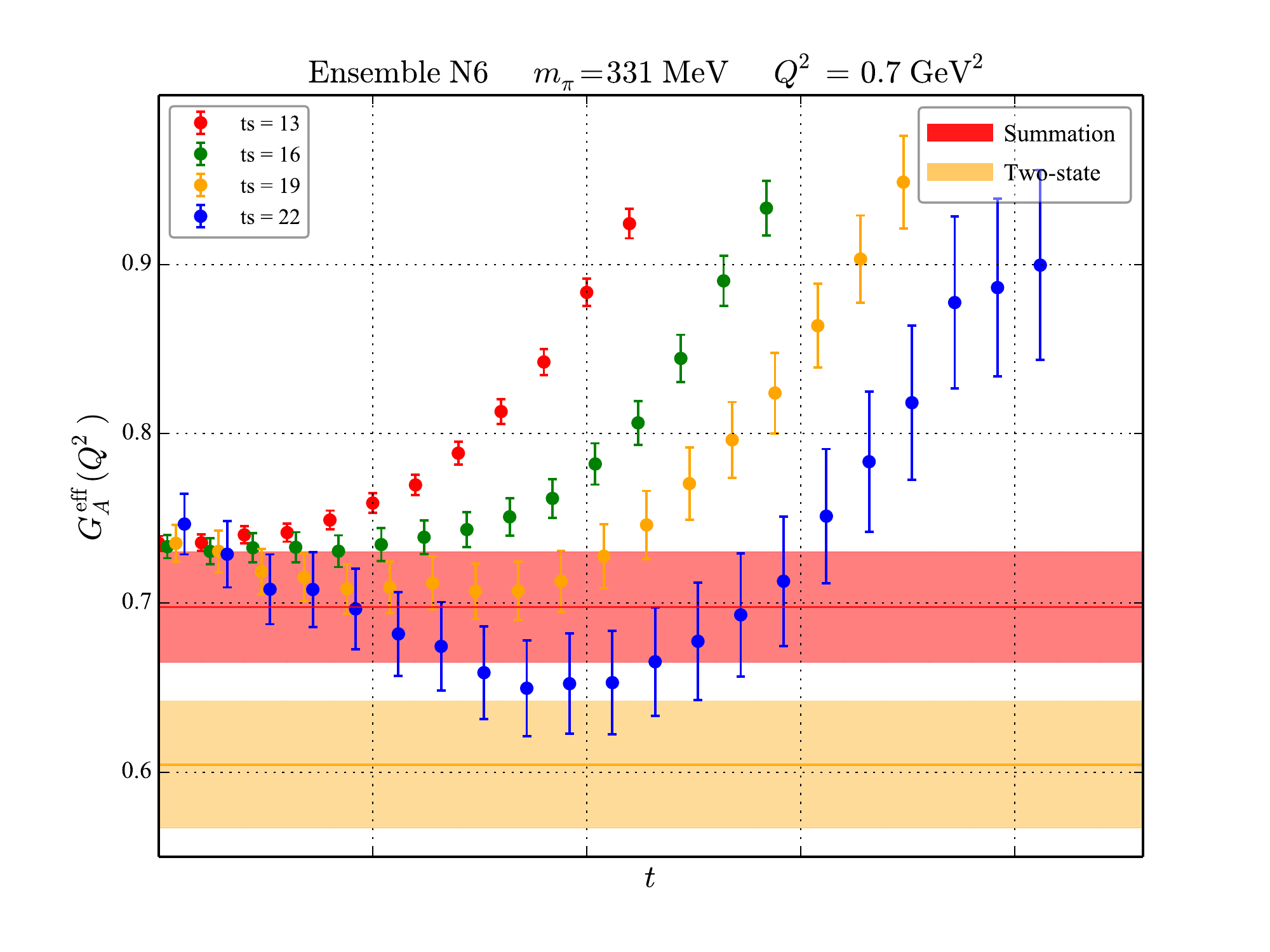}
        \caption{\label{Fig:effGA}
        The effective axial form factor $\GX{A}^{\rm eff}$ on the N6 ensemble
        at four-momentum transfers of $Q^2 = 0.3$~GeV$^2$ (left panel; corresponding to $(\frac{L \vec q}{2\pi})^2=1$)
        and $Q^2 = 0.7$~GeV$^2$ (right panel; corresponding to $(\frac{L \vec q}{2\pi})^2=3$). The bands represent the result
 for $\GX{A}(Q^2)$ obtained using the summation and the two-state-fit methods to extract asymptotic ratios,
after which the linear system resulting from Eq.\ (\ref{Eq:chisq}) is solved for $(\GX{A},\GX{P})$.}
\end{figure*}

In the left panel of Fig.~\ref{fig:gAN6}, results for the axial charge on
the N6 ensemble are shown, where the effective axial charge is computed for
four source-sink separations ranging from $t_s=0.6$~fm to $t_s=1.1$~fm.
The symmetric approach to the central plateau region can clearly be seen.
The data also exhibit a discrepancy between the midpoint values $g^{\rm eff}_{\rm A}(t_s/2,t_s)$ reached
at different source-sink separations, indicating that excited-state
contaminations are still present even when the ratio has apparently reached
a plateau. To investigate the excited-state contribution in the plateau
region further, the dataset was expanded to include source-sink separations
of $t_s=1.3$~fm and $t_s=1.4$~fm.
The right panel of Fig.~\ref{fig:gAN6} shows the results from applying a
plateau fit to the data at $t=t_s/2\pm 2$%
\footnote{For source-sink separations that are odd in lattice units,
plateaus are fitted at $(t_s-1)/2 \pm 2$}
for different source-sink separations $t_s$.
The dependence of the fit results on $t_s$ can be seen clearly. While
the large errors at the largest source-sink separations $t_s>1.1$~fm
somewhat obscure the trend, it is clear that $g_{\rm A}$ may be
underestimated when using plateau fits, and we do not employ plateau fits
in our further analysis.
Also shown is a fit to the expected $t_s$-dependence, taking the energy gap
to the excited state as a free parameter. The fit results are
compatible with the assumption of a dominant S-wave $N\pi\pi$ state,
although the uncertainty on the fit parameters is too large to make a
conclusive argument.

The results of various fit procedures are shown as coloured bands in
the left panel of Fig.~\ref{fig:gAN6}.  The blue band indicates a
plateau fit to source-sink separation $t=1.1$~fm, which is seen to lie
significantly below the results of both of the analysis methods used
in the following: the yellow band is a simultaneous fit to all
source-sink separations $t_s$ and operator insertion times $t$ up to
$1.1$~fm to the ansatz of Eq.~(\ref{Eq:gAansatz}), and the red band
indicates the result obtained using the summation method. The quality
of the linear fit performed in the latter method is illustrated in the
bottom panel of Fig.~\ref{fig:gAN6}.  The results from the two-state
fit and summation method agree very well with each other, indicating
that residual excited-state effects are likely small on this ensemble.
The effective axial charge on two of our most chiral ensembles is
displayed in Fig.~\ref{fig:gAG8F7}, together with the results of the
summation and excited-state fits shown as bands. On ensemble F7, the
two analysis methods are in very close agreement, while on G8 they
differ in their result for $g_{\rm A}$ by one standard deviation.

In Fig.~\ref{fig:gAextrapolation}, the results for $g_{\rm A}$ obtained
using the summation method (left panel) and two-state fits (right panel)
on each of our ensembles are shown together with a chiral extrapolation
to the physical pion mass.
Details of the chiral extrapolation procedure will be presented in
section~\ref{sec:chiral}. Here we note that the ensembles employed in
our work all obey the constraint $m_\pi L \geq 4$, and hence finite-volume
effects are expected to be small. This is also supported empirically by
our data, which do not display any noticeable volume dependence of $g_{\rm A}$.
Similarly, no discernable dependence of $g_{\rm A}$ on the lattice spacing is found in 
Fig.\ \ref{fig:gAextrapolation} for $m_\pi\leq 335\,{\rm MeV}$.

\begin{figure*}
        \centering
        \includegraphics[width=0.48\textwidth]{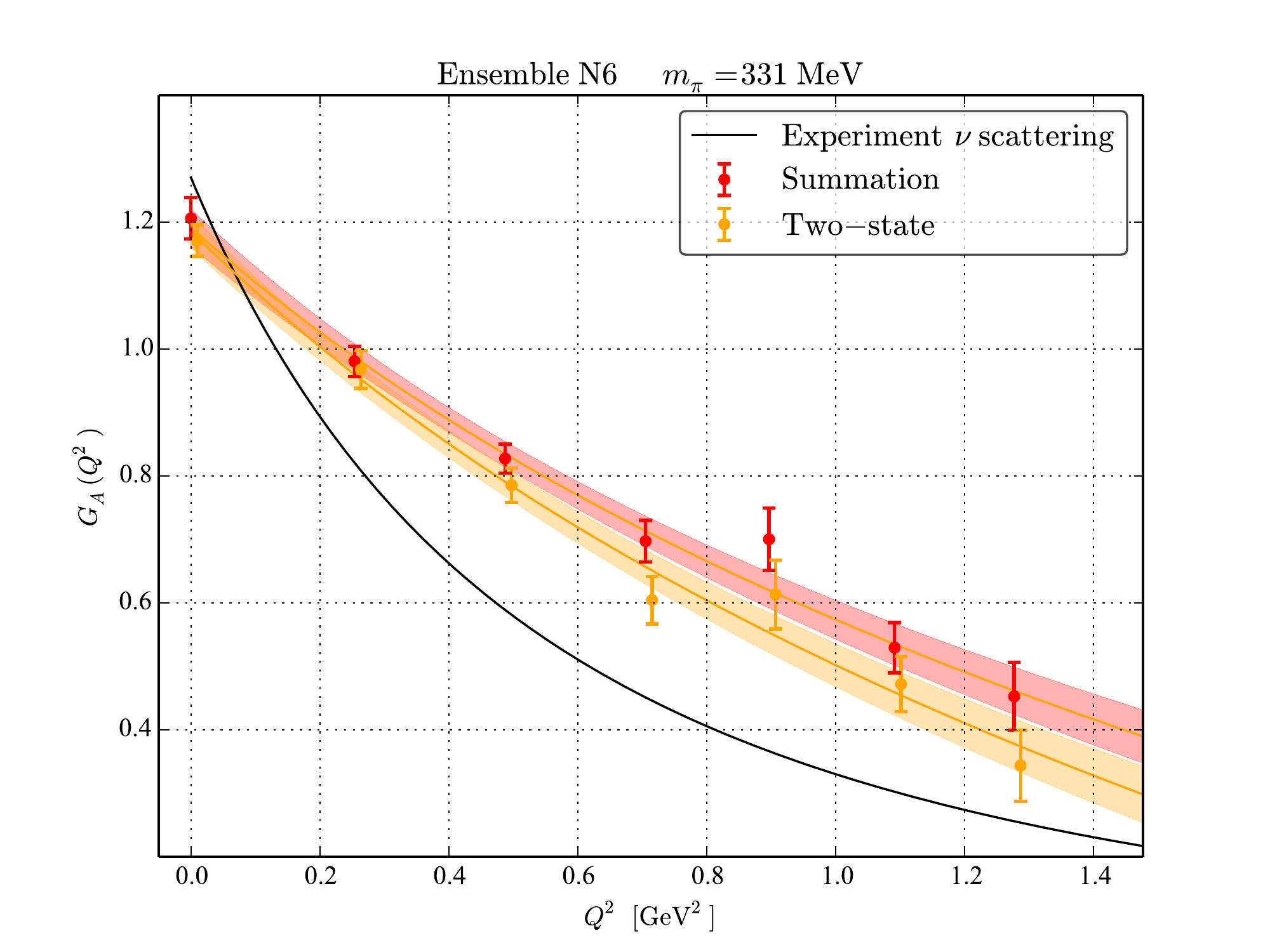}
        \includegraphics[width=0.48\textwidth]{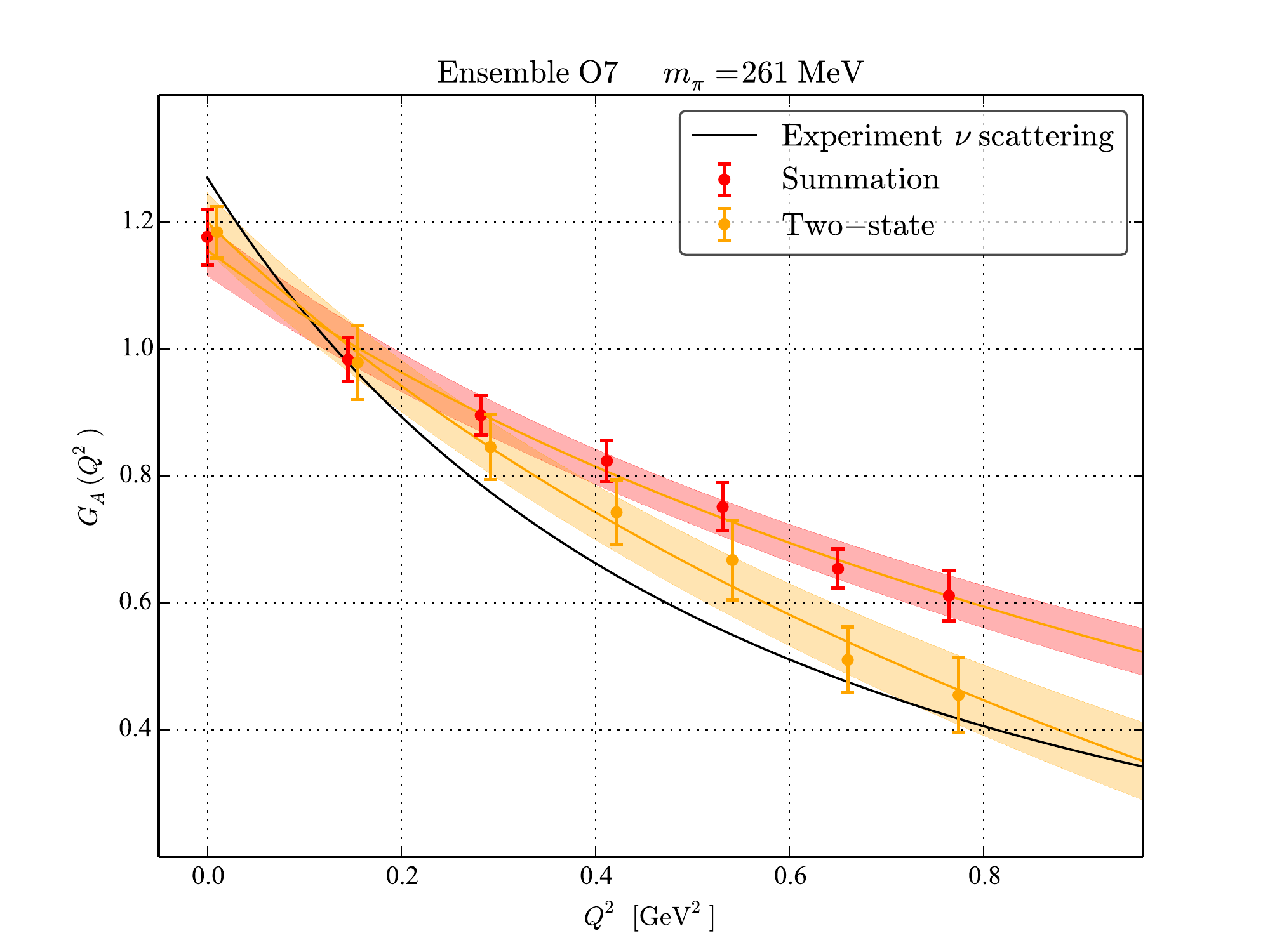}
        \caption{\label{Fig:qsqdep} Momentum-transfer dependence of the
                 axial form factor $\GX{A}(Q^2)$ on the ensembles
                 N6 (left panel) and O7 (right panel).
                 The solid black line shows a dipole parameterisation
                 of experimental data \cite{Bernard:2001rs}.}
\end{figure*} 

\begin{figure*}
        \centering
        \includegraphics[width=0.48\textwidth]{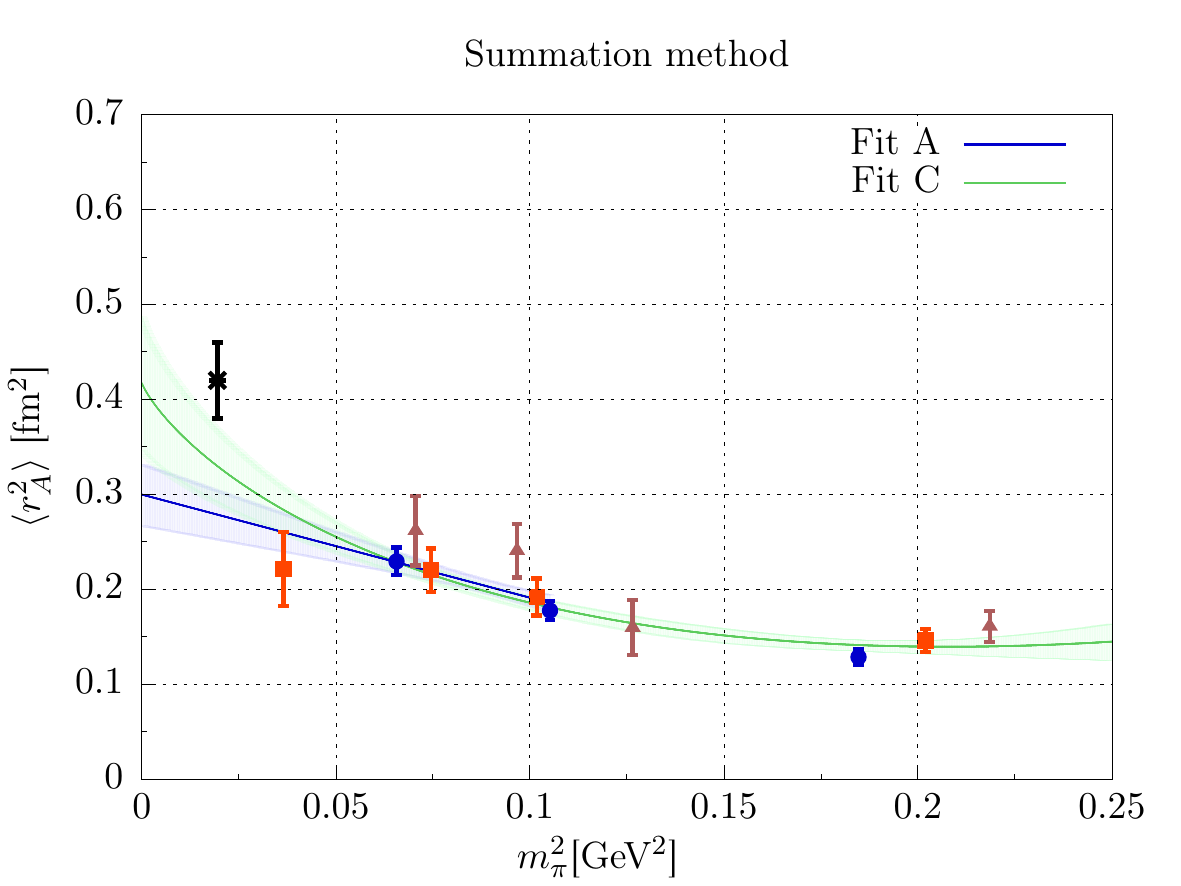}
        \includegraphics[width=0.48\textwidth]{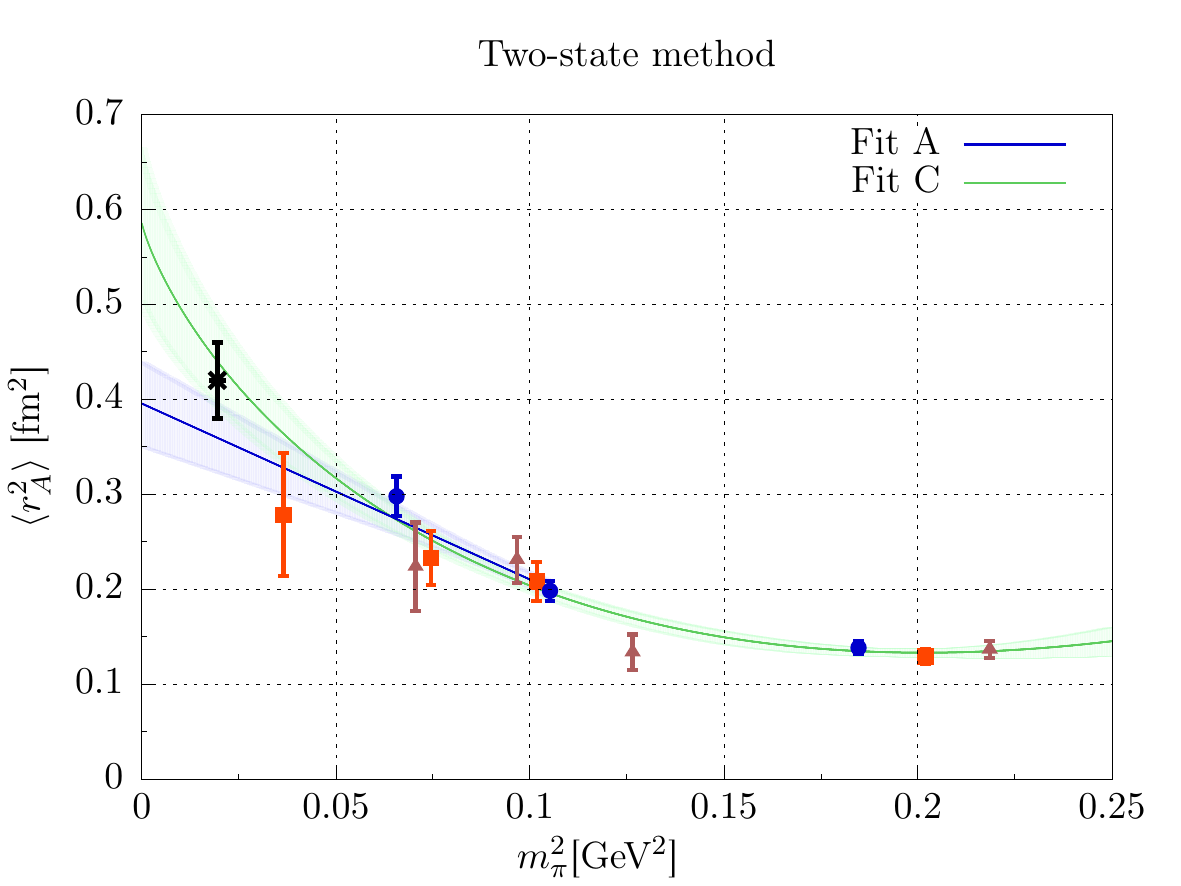}
        \caption{\label{Fig:rad_zexp} Results for the squared axial radius
                 $\langle r^2_{\rm A}\rangle$ from the $z$-expansion
                 applied to the summation method (left panel) and
                 two-state fit (right panel) results for $\GX{A}(Q^2)$.
                 Triangles, squares and circles correspond to increasingly fine
                 lattice spacings, and the black point represents the phenomenological
                 value of $\langle r^2_{\rm A}\rangle$~\cite{Ahrens:1988rr}. 
                 Fit A is a linear fit (\ref{eq:genfitlin}) with a pion-mass
                 cut $m_\pi\leq 335{\rm\,MeV}$, and 
                 Fit C is based on the ansatz (\ref{eq:genfitlog}) with no pion-mass cut.}
\end{figure*} 

\begin{table}[t]
\caption{\label{tab:sum_exc_chisq}Results for $\chi^2/{\text{dof}}$
         of the linear system in Eq.~(\ref{Eq:chisq})
         for each of the analysis methods used on Ensemble N6
         when including or excluding the pseudoscalar density
$P(x)$.\vspace{16pt}}
{
\renewcommand{\arraystretch}{2.0}
\begin{tabular}{c|cc|cc}\hline\hline
        \multirow{3}{*}{$\frac{\vec{q}^2L^2}{4 \pi^2}$} &
        \multicolumn{2}{c}{Summation Method} &
        \multicolumn{2}{c}{Two-state Method}\\ \cline{2-5}
         & $\chi^2/{\mathrm{dof}}$  & $\chi^2/{\mathrm{dof}}$ & $\chi^2 /{\mathrm{dof}}$ & $\chi^2/{\mathrm{dof}}$ \\   [-2ex]
         &  with $P(x)$  &  without $P(x)$  &  with $P(x)$  &  without $P(x)$ \\
         \hline
        1 & 90.0 &     n/a   & 1.85 &   n/a      \\ 
        2 & 23.3 & 0.41 & 0.13 & 0.26  \\ 
        3 & 19.8 &    n/a   & 0.07 &   n/a     \\ 
        4 & 15.2 &    n/a    & 1.88 &   n/a     \\ 
        5 & 3.09 & 1.59 & 1.03 & 0.21  \\ 
        6 & 1.76 & 0.23 & 1.10 & 0.02  \\ \hline\hline
\end{tabular}
}
\end{table}

\subsection{Momentum-transfer dependence of the axial form factor}

At non-zero momentum transfer, the determination of the axial form factor
becomes more involved. In our choice of reference frame, the $t$ dependence 
of the effective form factors is no longer symmetric about the point $t=t_s/2$;
as a consequence, we apply Eq.\ (\ref{eq:GenAnsatz}), as opposed to Eq.\ (\ref{Eq:gAansatz}), 
when using the two-state fit method.
Furthermore, both $\GX{A}$ and $\GX{P}$ contribute to the
matrix elements and must be separated by solving the linear system of
Eq.~(\ref{Eq:chisq}).
As described above in section \ref{sec:excited}, there are two ways
in which the solution of the linear system can be combined with the analysis
methods to account for excited-state contributions in order to extract the
ground-state form factors, and we shall employ both of these in the following.
Moreover, through the PCAC relation the matrix element of the pseudoscalar
density provides an additional observable that can be used in conjunction
with the matrix elements of the components of the axial current in order
to determine the axial-current form factors.
In the case of the axial form factor, we find that the determination of
the form factor $\GX{A}$ is relatively stable against the inclusion and
exclusion of the pseudoscalar density in our basis of operators.

The results of evaluating the effective axial form factor
$\GX{A}^{\rm eff}(Q^2,t,t_s)$ on the N6 ensemble at the four-momentum
transfers of $Q^2 = 0.3$~GeV$^2$ and $Q^2 = 0.7$~GeV$^2$ are shown
in Fig.~\ref{Fig:effGA}.
At $Q^2 = 0.3$~GeV$^2$, which is the lowest non-vanishing four-momentum
transfer that can be realised on this ensemble,
no clear plateau appears for source-sink separations in the range
of $t_s = 0.6-1.0$~fm, and for the largest source-sink separation
$t_s = 1.1$~fm, the size of the uncertainties on the data make it difficult
to decide whether a plateau has truly been reached. 

Also shown are bands indicating the results of fitting the effective form
factor using a two-state fit (Eq.\ (\ref{eq:GenAnsatz})) and the summation method (Eq.\ (\ref{eq:summation})), 
and it can be seen that these agree with each other within their respective uncertainties.
At the larger four-momentum transfer of $Q^2 = 0.7$~GeV$^2$, on the other
hand, the results for the effective form factor $\GX{A}^{\rm eff}(Q^2,t,t_s)$
at different source-sink separations $t_s$ show a clear downward trend in the central region, $t\approx t_s/2$, where excited state contributions are expected to be most
strongly suppressed, indicating that the plateau has not stabilized
for $t_s\gtrsim 1.1$ fm. Furthermore, the results for $G_A(Q^2)$
obtained using the summation method and two-state fits are not in
agreement with each other, which may be due to the more statistically
precise data at the lower source-sink separations having a
disproportionately strong influence on the fit results. This may
introduce a bias in the two-state fit, but is likely to affect the
slope of the summed ratio in the summation method even more.
We will return to considering the
relative reliability of the two methods when discussing the induced
pseudoscalar form factor.

To confirm the stability of our analysis, we have verified that we obtain
the same results for the axial form factor $\GX{A}(Q^2)$ at each four-momentum
transfer $Q^2$ when we fit to the asymptotic behaviour of the effective
form factors, and when we first extract the asymptotic behaviour of the ratios
before isolating the form factors (i.e. methods 1 and 2 of section \ref{sec:excited}
yield consistent results).
The values of $\chi^2/\text{dof}$ obtained when solving the linear system
of Eq.~(\ref{Eq:chisq}) after extracting asymptotic ratios using the summation
method or two-state fits are shown in Table \ref{tab:sum_exc_chisq}.
The values of $\chi^2/\text{dof}$ obtained with the
summation method are large when $P(x)$ is included.
Nevertheless the results for the axial form factor are quite stable, regardless of
whether the pseudoscalar density $P(x)$ is excluded or included in the analysis.
This may be attributed largely to the fact that the inclusion of $P(x)$
affects the results for the induced pseudoscalar form factor $\GX{P}(Q^2)$
much more strongly than those for $\GX{A}(Q^2)$; we will further remark on
this when discussing the induced pseudoscalar form factor in
section~\ref{sec:GP}.
When extracting the asymptotic ratios using two-state fits, the values
of $\chi^2/\text{dof}$ for Eq.~(\ref{Eq:chisq}) are rather reasonable
both when including and when excluding the pseudoscalar density,
and the results for $\GX{A}(Q^2)$ (and for $\GX{P}(Q^2)$, 
as discussed in section~\ref{sec:GP}) are likewise compatible.

The momentum-transfer dependence of the axial form factor is shown in
Fig.~\ref{Fig:qsqdep} for the ensembles N6 ($m_\pi = 331 $~MeV)
and O7 ($m_\pi = 261$~MeV).
We note that while the results from the summation method and from
two-state fits agree well near $Q^2=0$, the disagreement between them
becomes larger with increasing $Q^2$, where the two-state fits tend to
approach more closely the shape of the form factor inferred from
experimental results, and this trend becomes more pronounced as the pion
mass is decreased towards the physical point.
We also note that (as previously observed 
elsewhere\cite{Yamazaki:2008py,Green:2012ud,Capitani:2012gj,Horsley:2013ayv,Bhattacharya:2013ehc,
Bali:2014nma,Chambers:2014qaa,Bhattacharya:2016zcn,Bouchard:2016heu,Yoon:2016jzj,Liang:2016fgy,Berkowitz:2017gql})
while the lattice calculations tend to underestimate
the axial charge, they tend to overestimate the value of the form factor
at non-vanishing four-momentum transfer, leading to an underestimation of
the axial radius of the nucleon.

\subsection{Model-independent determination of $\langle r^2_{\rm A} \rangle$}

While the momentum-transfer dependence of the axial form factor is
frequently modelled with a dipole fit \cite{Bernard:2001rs}, this leads
to a model-dependence of the determination of the axial charge radius
$\langle r^2_{\rm A} \rangle^{1/2}$. Moreover, the use of the momentum transfer
$Q^2$ as the expansion variable has been shown to have a small radius
of convergence, and the use of a conformally mapped parameter $z(Q^2)$ has
been suggested \cite{Hill:2010yb,Bhattacharya:2011ah} in order to improve
the convergence by parameterising the form factor in a model-independent
manner as a power series in $z(Q^2)$.
The definition of  $z(Q^2)$ and the corresponding power series for the
form factor are given by
\begin{equation}\label{Eq:zexp}
z(Q^2)  = \frac{\sqrt{t_{\rm cut} + Q^2} - \sqrt{t_{\rm cut}} }{\sqrt{t_{\rm cut} + Q^2} + \sqrt{t_{\rm cut}}}, \quad G_{\rm A}(Q^2) = \sum^{\infty}_{n=0} a_n z^n(Q^2) \,,
\end{equation}
where $t_{\rm cut} = 9 m^2_\pi$ is the three-pion kinematic threshold in the
iso-vector axial-current channel. 
The power-series expansion of the form factor shown in Eq.~(\ref{Eq:zexp})
provides a controlled way of obtaining observables such as the axial radius
in a model-independent fashion: once the coefficients $a_n$ have been
determined from a fit to Eq.~(\ref{Eq:zexp}), the axial radius as defined
in Eq.~(\ref{Eq:radius}) can be derived from them in a straightforward manner.

\begin{figure*}[t]
        \centering
        \includegraphics[width=0.48\textwidth]{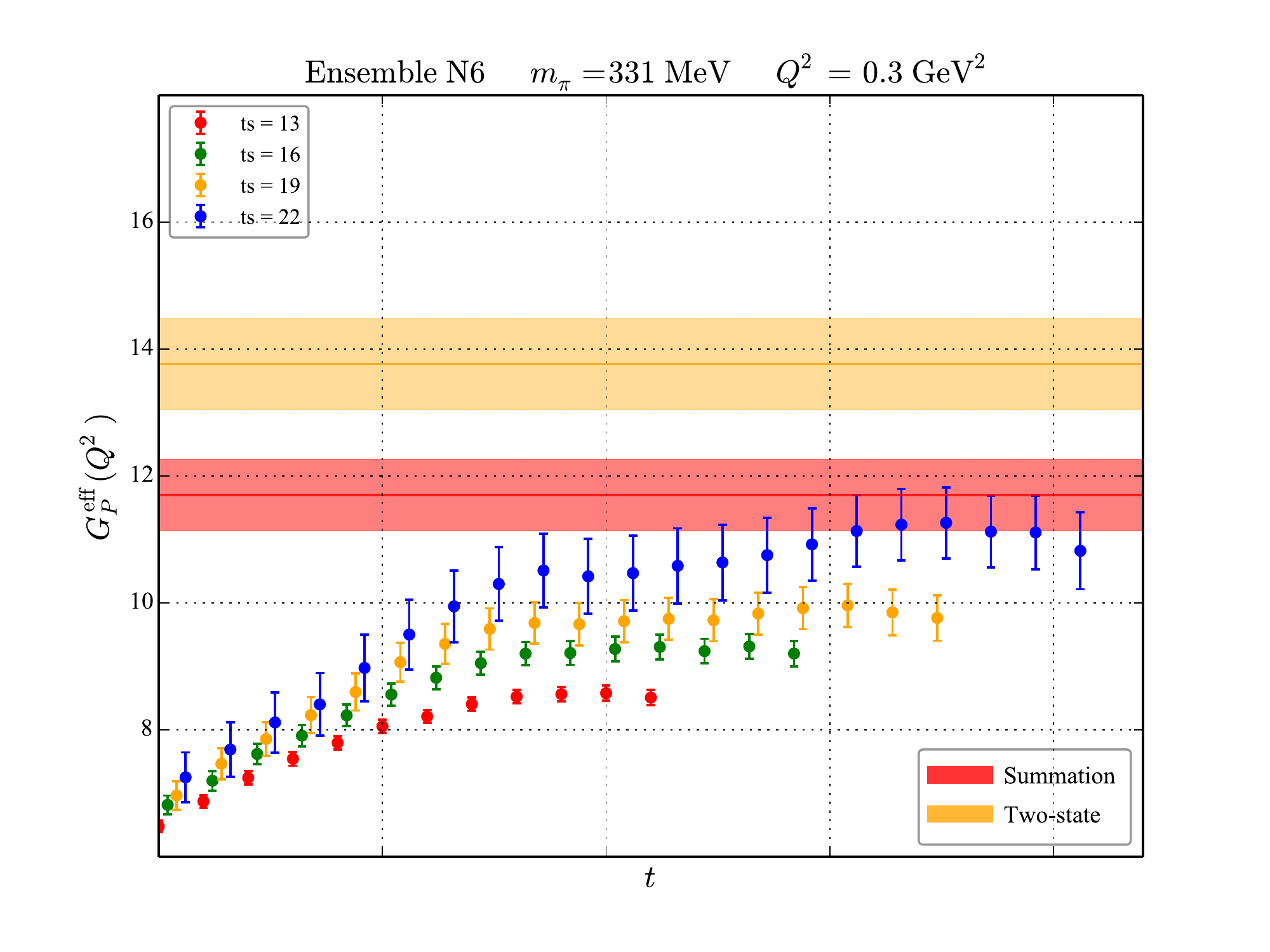}
        \includegraphics[width=0.48\textwidth]{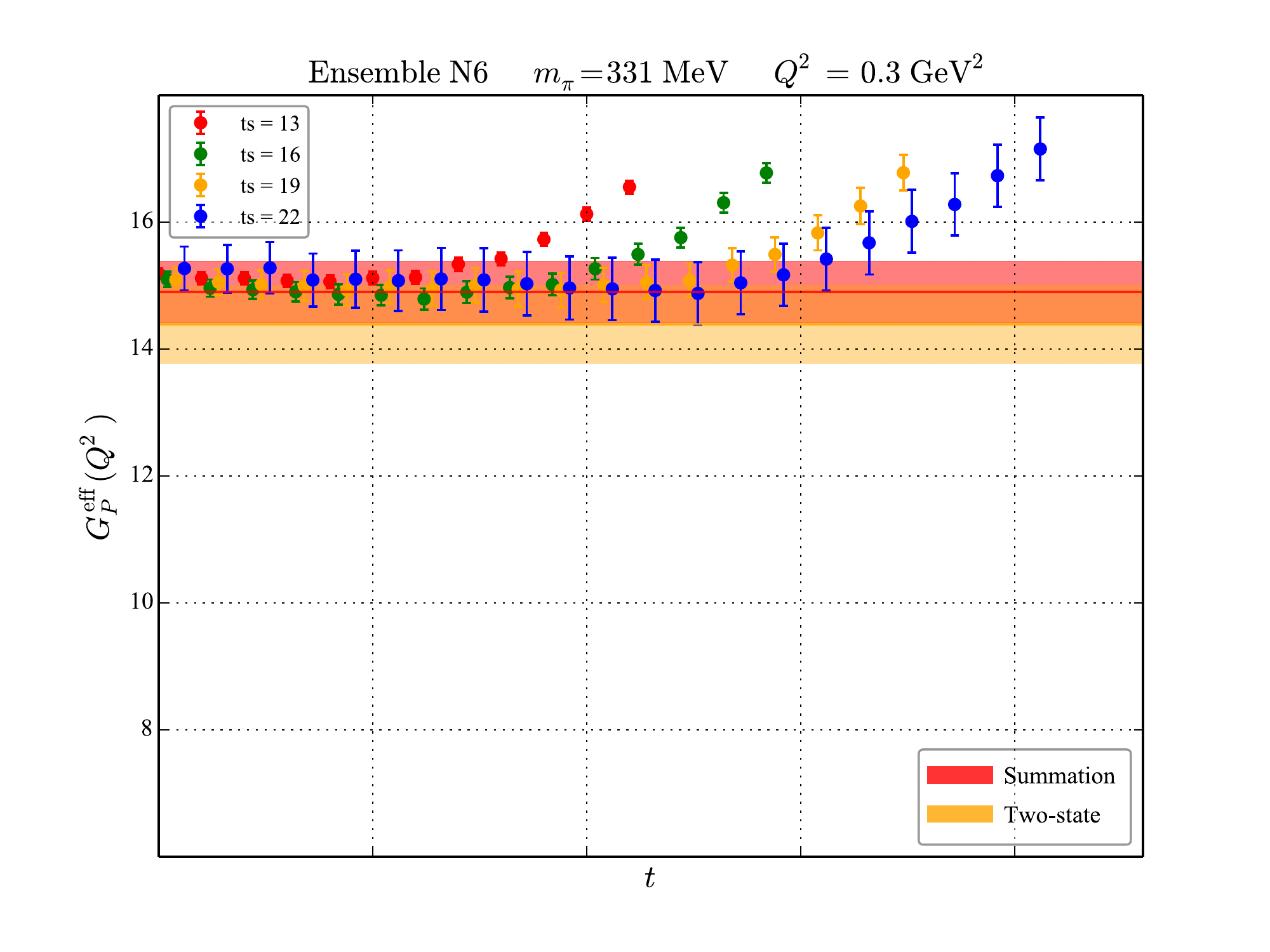}
        \caption{\label{Fig:effGP}
                 Effective induced pseudoscalar form factor on the N6 ensemble
                 at four-momentum transfer $Q^2 = 0.3$ GeV$^2$ when
                 excluding (left panel) and including (right panel) the
                 pseudoscalar density $P(x)$ in the basis of ratios used
                 to extract the form factors. The bands represent the result
                 for $\GX{P}(Q^2)$ obtained using the summation and the two-state-fit
                 methods to extract asymptotic ratios,
                 after which the linear system resulting from Eq.\ (\ref{Eq:chisq}) is
                 solved for $(\GX{A},\GX{P})$.}
\end{figure*}

\begin{figure*}[t]
        \centering
      \includegraphics[width=0.48\textwidth]{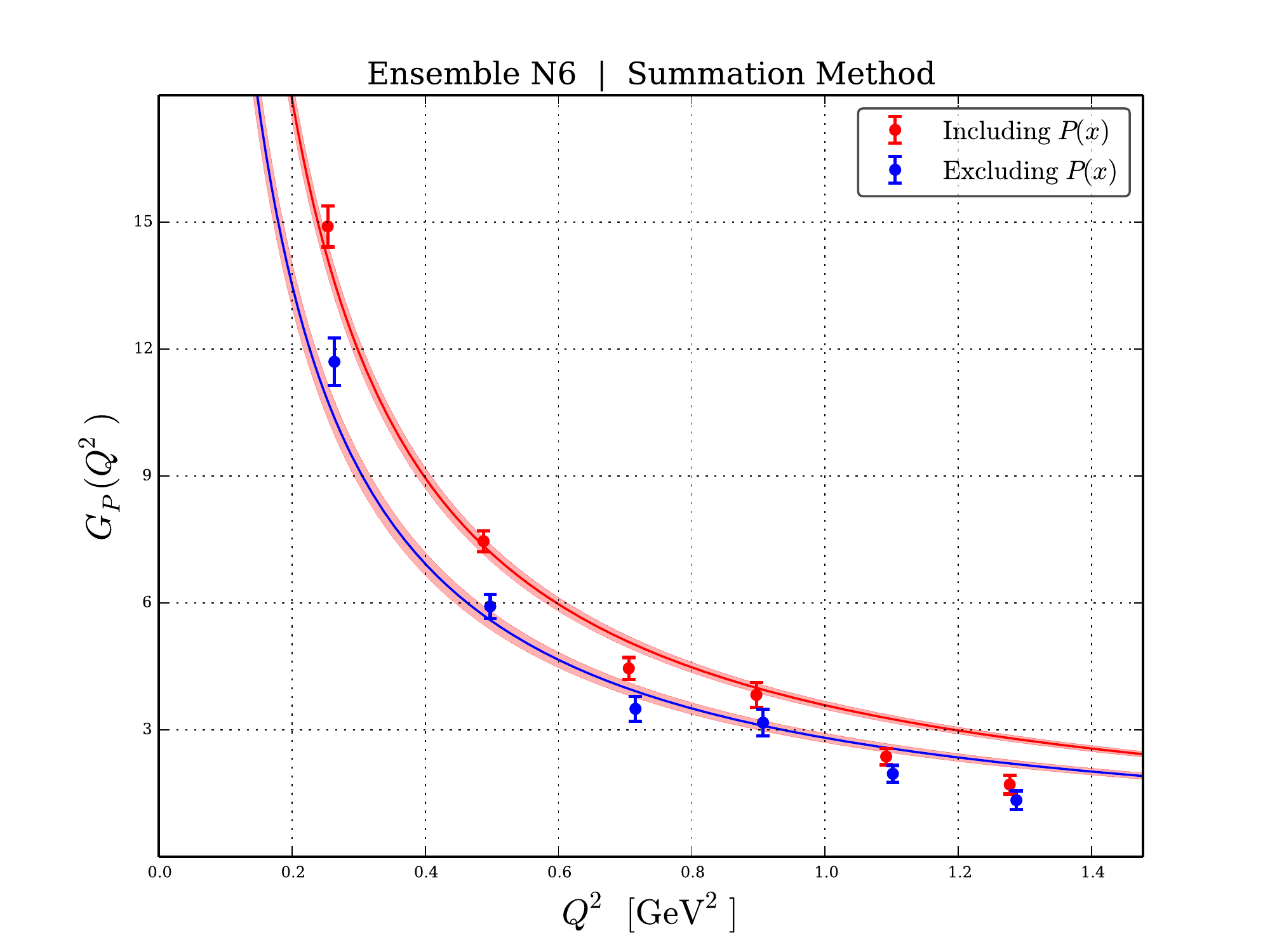}
        \includegraphics[width=0.48\textwidth]{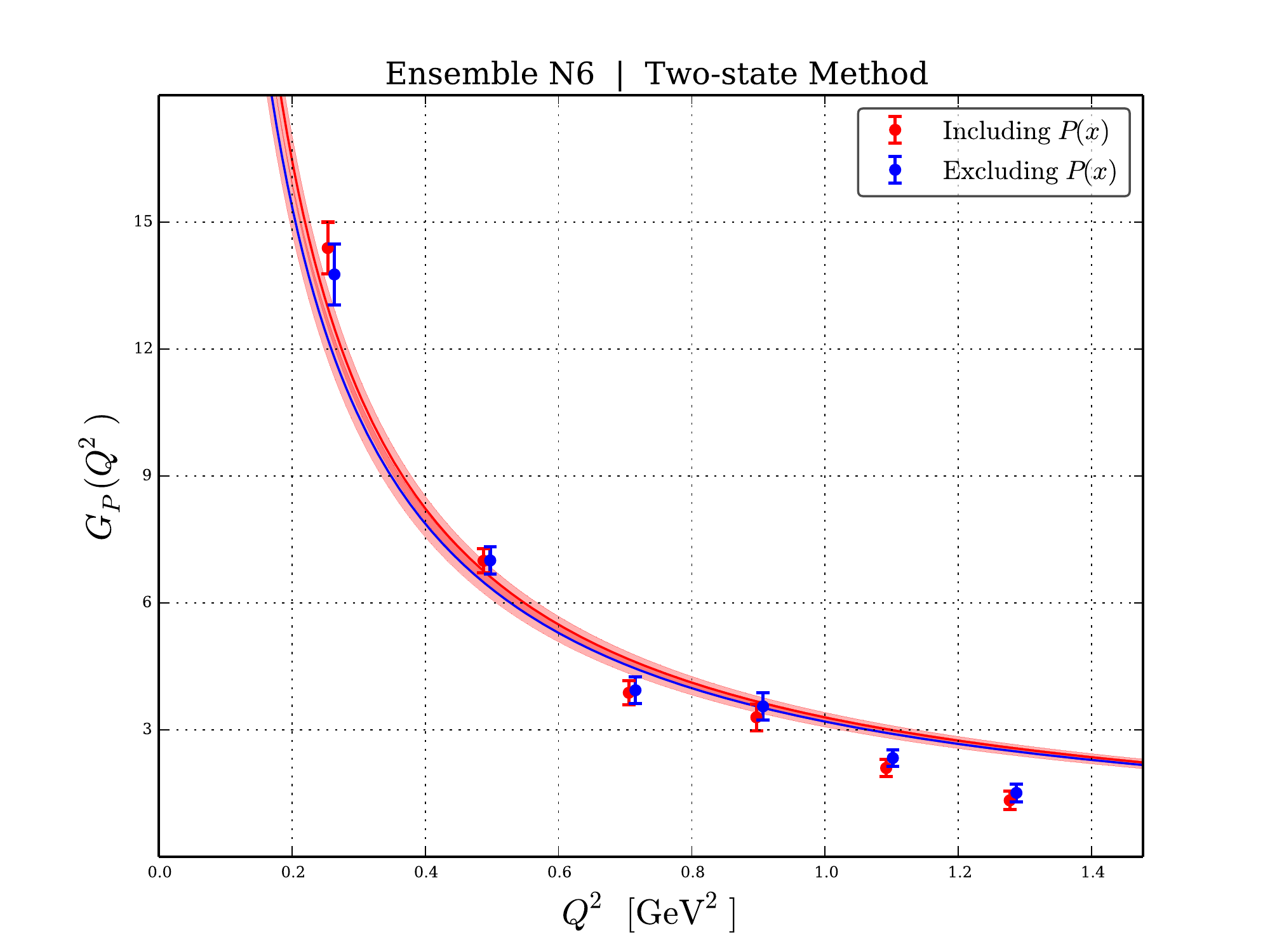}
        \caption{\label{Fig:GP_comparison}
                 Results for the induced pseudoscalar form factor obtained
                 by solving Eq.~(\ref{Eq:chisq}) for the asymptotic ratios
                 on the N6 ensemble using the summation method (left panel)
                 and two-state fits (right panel). The curves represent a fit using
                 the Goldberger-Treiman-inspired pion-pole parameterization
                 $\GX{P}(Q^2) = 4 M^2_N \GX{A}(Q^2)/(Q^2 + m^2_\pi)$,
                 where $M_N$ and $m_\pi$ are obtained from the fit.}
\end{figure*}

In this work, we have studied up to $n_{\rm max}=4$ orders in the
$z$-expansion of Eq.~(\ref{Eq:zexp}), and the results at different
orders were found to be consistent, provided that Bayesian priors were
used to stabilize the fit; otherwise, the fits beyond $n_{\rm max}=1$
became too unstable.  While we have checked that the results for the
axial charge radius obtained from the $z$-expansion were stable
against variations of the priors, we quote only the results obtained
using the first order of the $z$-expansion, where no priors were
applied.  The results obtained for the axial charge radius on our set
of ensembles using the summation method and two-state fits are given in Tab.~\ref{tab:rAgp} and 
presented in Fig.~\ref{Fig:rad_zexp} together with a chiral
extrapolation to the physical pion mass. For details of the chiral
extrapolation, the reader is referred to section \ref{sec:chiral}.
For now, we make the qualitative observation that the mean-square
radius increases by roughly a factor two between $m_\pi=430\,{\rm
  MeV}$ and $m_\pi=270\,{\rm MeV}$. Similar to the axial charge, no
systematic trend of the axial radius as a function fo the lattice
spacing is seen.


\section{Isovector induced pseudoscalar form factor}
\label{sec:GP}

The momentum transfer dependence of the induced pseudoscalar form factor
$\GX{P}(Q^2)$ is markedly different from that of the axial form factor
due to the presence of a pole at the pion mass arising as a consequence
of chiral symmetry breaking.
The low-momentum behaviour of $\GX{P}(Q^2)$ may therefore be expected to
be rather steep, with possibly considerable statistical fluctuations in the
low-$Q^2$ region.
In this section, we discuss the determination of the induced pseudoscalar
form factor from our various analysis methods for suppressing
excited-state contributions, and how these are affected by the choice of either
including or excluding the pseudoscalar density in our operator basis.

Relation (\ref{Eq:pcac_mat}) expresses the PCAC relation at the level
of ground-state matrix elements.  Since the extraction of the latter
from the three-point function comes with an additional uncertainty due
to potential contamination from excited states, we initially test the
PCAC relation at the correlator level,
\begin{eqnarray}
&& \frac{R_{A_0}(\vec q,t,t_s)}{C_{3,A_0}(\vec q,t,t_s)}\,\frac{1}{2a}\Big( C_{3,A_0}(\vec q,t+a,t_s)- C_{3,A_0}(\vec q,t-a,t_s) \Big)
\\ &&  - i {\textstyle\sum_{k=1}^3} q_k R_{A_k}(\vec q,t,t_s) - 2m_q R_P(\vec q,t,t_s) = {\rm O}(a^2).
\nonumber
\end{eqnarray}
We have checked that the left-hand side has an average compatible with
zero. For instance, on ensemble O7 at the smallest available spatial
momentum, the absolute statistical uncertainty on the left-hand side
is no more than 0.01 for all values of $t$ and
$t_s$~\cite{HuaThesis}. Thus the violation of the PCAC relation due to
discretization errors appears to be small. Therefore, we adopt the
point of view that the consistency condition (\ref{Eq:pcac_mat}) on
the ground-state matrix elements can be used as a way to test the ability 
of the summation method and the two-state fits to extract 
the ground-state matrix elements.

In order to gain insight into the nature of the excited-state contributions,
we first construct the effective form factor $\GX{P}^{\rm eff}(Q^2,t,t_s)$
by solving the linear system of Eq.~(\ref{Eq:chisq}) at each four-momentum
$Q^2$, operator insertion time $t$, and source-sink separation $t_s$. 
In doing so, we find a marked difference in the time-separation dependence of
$\GX{P}^{\rm eff}(Q^2,t,t_s)$ between the case where only components of
the axial current are included in solving Eq.~(\ref{Eq:chisq}) and the case
where a combination of components of the axial current $A_i(x)$ as well as
pseudoscalar density $P(x)$ are included. This is in contrast to the situation
for the axial form factor $\GX{A}(Q^2)$, where no strong dependence on the
inclusion or exclusion of $P(x)$ in the operator basis was observed.
The results for both cases are shown in Fig.~\ref{Fig:effGP}.
It can be clearly seen that the choice of whether to include the pseudoscalar
density $P(x)$ has a significant and non-trivial impact on the time-dependence
of the effective induced pseudoscalar form factor.
For the case where only the axial current is included in the determination
(shown in the left panel of Fig.~\ref{Fig:effGP}), the results asymptote
from below, with much stronger excited-state effects visible at the source
as compared to the sink. The apparent plateaux reached at different source-sink
separations $t_s$ do not agree with each other, indicating that the
contribution from excited states is substantial at time separations as large
as $1.1$~fm.
For the case where both the pseudoscalar density and the axial current are
included in the determination (shown in the right panel of
Fig.~\ref{Fig:effGP}), the results are much less time-dependent and asymptote
from above, with stronger excited-state effects seen at the sink rather than
the source. The plateaux for different source-sink separations $t_s$ agree
with each other within their respective statistical uncertainties.
However, the plateau values differ significantly from those seen when
excluding $P(x)$, even at the largest source-sink separations, which further
indicates that excited-state contamination remains a significant effect even
at $t_s\sim 1.1$~fm.

Also shown in Fig.~\ref{Fig:effGP} are the results of applying each of
our excited-state analysis methods, {\em viz.} the summation method (red bands)
and two-state fits (yellow bands). It can be seen that when extracting
the induced pseudoscalar form factor using only the axial current, the
results from the two methods disagree significantly, which may indicate that
excited-state effects are not under control.
When including the pseudoscalar density in the determination, on the other
hand, the results from the summation and two-state methods agree within their
respective error bands.
As pointed out in section \ref{sec:GA}, the matrix elements of the
pseudoscalar density are found to be statistically more precise in comparison
to those of the axial current, and hence strongly influence the determination of the
effective form factors. Their impact on the results seems to be limited to
the induced pseudoscalar form factor, however.
Judging both from the appearance of the effective form factor
and from the agreement between our analysis methods, the excited-state effects
seem to be smaller in the case where the pseudoscalar density is included.
We conclude that its inclusion is beneficial and therefore keep 
the pseudoscalar density as part of all subsequent analyses.

The dependence on $Q^2$ of the results obtained using each of our analysis
methods when including or excluding the pseudoscalar density $P(x)$ in the
solution of Eq.~(\ref{Eq:chisq}) is presented in Fig.~\ref{Fig:GP_comparison}.
The form factors obtained with the summation method (shown in the left panel of
Fig.~\ref{Fig:GP_comparison}) can be seen to be particularly sensitive to
the inclusion of the pseudoscalar density, with a clear gap opening up
particularly in the low-$Q^2$ regime.
This is also evident from the large values of $\chi^2/\text{dof}$ for the
solution of Eq.~(\ref{Eq:chisq}) shown in the relevant columns of
Table~\ref{tab:sum_exc_chisq} and may indicate that the summation method
is not able to properly account for the large excited-state contamination
found in $\GX{P}(Q^2)$ when using only the axial current.
By contrast, the results obtained using the two-state method (shown in the
right panel of Fig.~\ref{Fig:GP_comparison}) indicate good stability
against the choice of including or excluding the pseudoscalar density.
This observation is corroborated by the values of  $\chi^2/\text{dof}$ shown in the
relevant columns of Table~\ref{tab:sum_exc_chisq}.
It is also worth noting that while the results from the summation method
show a high sensitivity to the choice of the operator basis used in solving
Eq.~(\ref{Eq:chisq}), a marked improvement in the compatibility with the
results from the two-state method is observed when the pseudoscalar
density is included.
Since judging from their stability under the choice of operator basis,
the results obtained with the two-state method appear to be more reliable,
the two-state method will be the method of choice in our subsequent analysis.

In summary, the ground-state matrix elements extracted with the
two-state fit method are overall consistent with the PCAC relation
Eq.\ (\ref{Eq:pcac_mat}), which is a non-trivial check that
excited-state contaminations have been removed, since the extraction
of each term in Eq.\ (\ref{Eq:pcac_mat}) may be affected differently
by excited states. By contrast, the ground-state matrix elements
extracted with the summation method do not satisfy the PCAC relation, as 
seen from the large reduced $\chi^2$ values in Table~\ref{tab:sum_exc_chisq};
we therefore do not use the summation method for our final results,
but nonetheless quote intermediate results derived from it 
in tables \ref{tab:bigfit} and \ref{tab:chiralfits}.
In the future, we hope to carry out high-statistics calculations 
at such large source-sink separations that both methods yield
ground-state matrix elements that are consistent with each other
and with the PCAC relation.


\section{Chiral Analysis}
\label{sec:chiral}

In order to provide predictions for the physical world, it remains
to extrapolate our lattice results obtained at unphysical values of the
light quark masses to the physical pion mass and the continuum limit.
The standard approach to this problem is to take the values of the
observables of interest ($g_{\rm A}$ and the axial charge radius in our case)
as determined on each ensemble, and to extrapolate them to the physical
point using formulae taken from, or inspired by, Chiral Perturbation
Theory~(ChPT).

\subsection{Combined fit in chiral effective theory}

\begin{table}[t]
\caption{The low-energy constants relevant to the EFT description
of the nucleon axial-vector form factors \cite{Schindler:2006it},
with the values used in the fits to our lattice data. Quantities with
a circle on top denote values in the chiral limit; in the case of the nucleon
mass, higher-order effects are accounted for by using either the physical nucleon
mass or the measured value on each lattice ensemble.\vspace{16pt}}
{
\label{tab:lecs}
\renewcommand\arraystretch{1.5}
\begin{tabular}{ccl}\hline\hline
Interaction & Low-energy constant & Value and role in the fit\\\hline
$\mathcal{L}^{(2)}$ & $F$         & $F_\pi^{\rm exp} =92.2$~MeV \\
                    & $m_\pi^2$   & Lattice input \\\hline
$\mathcal{L}_A^{\rm eff}$ & $M_A$ & $M_{a_1}=1.23$~GeV (Ref.
\cite{Olive:2016xmw}) \\\hline
$\mathcal{L}^{(1)}_{\pi N}$ & $F_G$       & Fit parameter contributing to $\langle r_{\rm A}^2\rangle$ and $\GX{A,P}$ \\
                    & $\mathring{m}_N$ & Lattice input \\ && or $m_N^{\rm phys}=938.3$~MeV\\
                    & $\mathring{g}_{\rm A}$ & Fit parameter contributing to $\GX{A}$ and $\GX{P}$ \\\hline
$\mathcal{L}^{(2)}_{\pi N}$ & $c_3$       & $c_3 = -4.2/m_N^{\rm phys}$ (Ref.
\cite{Becher:2001hv})\\
                    &                     & or $c_3 = -5.61$~GeV$^{-1}$ (Refs.
\cite{Hoferichter:2015hva,Hoferichter:2015tha})\\
                    & $c_4$       & $c_4 = 2.3/m_N^{\rm phys}$ (Ref.
\cite{Becher:2001hv})\\
                    &             & or $c_4 = 4.26$~GeV$^{-1}$ (Refs.
\cite{Hoferichter:2015hva,Hoferichter:2015tha})\\\hline
$\mathcal{L}^{(3)}_{\pi N}$ & $d_{16}$    & Fit parameter contributing to $\GX{A}$ and $\GX{P}$\\
                    & $d_{18}$    & Fit parameter contributing to $\GX{P}$ \\
                    & $d_{22}$    & Fit parameter contributing to $ \langle r_{\rm A}^2\rangle$ and $\GX{A,P}$ \\ 
\hline\hline
\end{tabular}
}
\end{table}

One possible approach that can be applied here is very similar to that
performed in our paper \cite{Capitani:2015sba} on the electromagnetic
form factors of the nucleon:
we perform a fit of the dependence of the form factors $\GX{A}$ and $\GX{P}$
on both the pion mass and the squared momentum transfer $Q^2$ to the
expressions of baryonic effective field theory~(EFT), including 
explicit axial vector degrees of freedom \cite{Schindler:2006it}.
The main motivation for this ansatz is that the inclusion of the axial
vector meson extends the range in $Q^2$ for which a phenomenologically
good description of the form factor data is achieved \cite{Schindler:2006it}. This
increases the number of points amenable to a simultaneous fit to the
$Q^2$ and pion mass dependence of the form factors $G_A$ and
$G_P$. From such a fit we extract the LECs and subsequently use the
so obtained values at the physical pion mass to quote the extrapolated
result for our data. This approach avoids the two-step procedure 
of first extracting the form factors and derived
quantitites using a dipole fit or a $z$-expansion ana\-ly\-sis, before
applying a chiral extrapolation. It also has the potential advantage that 
values are obtained for low-energy constants that can be used in describing
other obervables.

The full analytic form of the ansatz is quite lengthy; it can be found in
\cite{Schindler:2006it}. 
While we do not reproduce it in its full length here, 
the contributions of the fit parameters to $\GX{A}(q^2)$ and $\GX{P}(q^2)$ are simple to write down, 
\begin{eqnarray}
G^{\rm EFT}_{\rm A}(q^2)&=&\stackrel{\circ}{g}_A+4m_\pi^2d_{16} -d_{22} q^2 - 8 F_G 
\frac{q^2}{q^2-M_{A}^2}
+\mathcal{O}(\hbar),\\
G^{\rm EFT}_{\rm P}(q^2)&=& 4m_N^2 d_{22} +\frac{4m _N^2  }{q^2-m_\pi^2} 
\left(-\stackrel{\circ}{g}_A
+ 2 m_\pi^2 (d_{18}-2 d_{16})  +\mathcal{O}(\hbar) \right)
\nonumber\\ && +32 \frac{F_G m_N^2}{q^2-M_{A}^2}  + \mathcal{O}(\hbar),
\end{eqnarray}
where $\mathcal{O}(\hbar)$ indicates terms of at least one-loop order.
The fitted low-energy constants are
$\mathring{g}_{\rm A}$, $d_{16}$, $d_{18}$, $d_{22}$ and $F_G$,
as summarized in Table \ref{tab:lecs}. The low-energy constants
$c_3$ and $c_4$, and the mass $M_{A}$ of the axial vector meson 
are set to their phenomenological values \cite{Becher:2001hv},
while for the nucleon mass its measured value on each ensemble is used.
A pion-mass cut of $m_\pi\le 280$~MeV is applied to the fits, which thereby only
include four ensembles, as well as a momentum cut of $Q^2\le 0.4$~GeV$^2$.
The reason for the reduction in the $m_\pi$ fit range 
is that we obtained values for the $\chi^2/{\rm d.o.f}$ of about 4-5 with
the less restrictive choice of $m_\pi<335\,$MeV made in the next subsection.
We perform a simultaneous fit to both $\GX{A}(Q^2)$ and $\GX{P}(Q^2)$
using a fit function accounting for the leading discretisation effects,
\begin{equation}
\GX{A,P}(Q^2) = \GX{A,P}^{\rm EFT}(Q^2,m_\pi) + e_1^{A,P} a^2 + e_2^{A,P} Q^2a^2.
\end{equation}
In order to estimate the influence of various systematic effects, we also
perform a number of variations on the fit by
\begin{enumerate}
\item neglecting discretisation effects ($e_1^{A,P}=e_2^{A,P}=0$),
\item not applying a cut in $Q^2$,
\item using the physical nucleon mass on all ensembles, and
\item using different values for the low-energy constants $c_3$ and $c_4$.
\end{enumerate}

\begin{figure}
\begin{center}
\includegraphics[width=0.65\textwidth,keepaspectratio=]{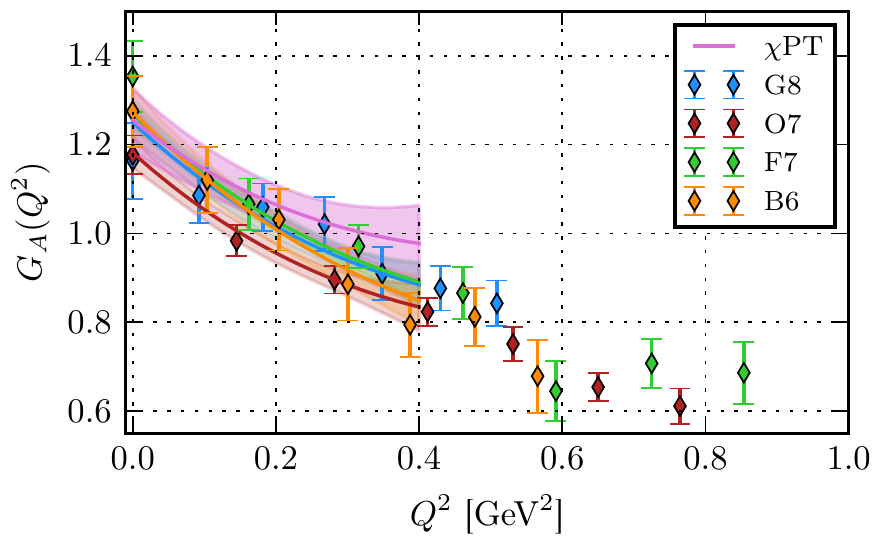}
\end{center}
\caption{Our results for the axial form factor $\GX{A}(Q^2)$ of the nucleon,
with the chiral fit and its extrapolation to the physical point.
The band in the same colour as the data point for each ensemble represents the result of
inserting into the fit function the pion mass and lattice spacing of that ensemble,
and the purple band labelled ``$\chi$PT'' represents the fit function in the continuum
limit at the physical pion mass.
}
\label{fig:GAbigfit}
\end{figure}

\begin{figure}
\begin{center}
\includegraphics[width=0.65\textwidth,keepaspectratio=]{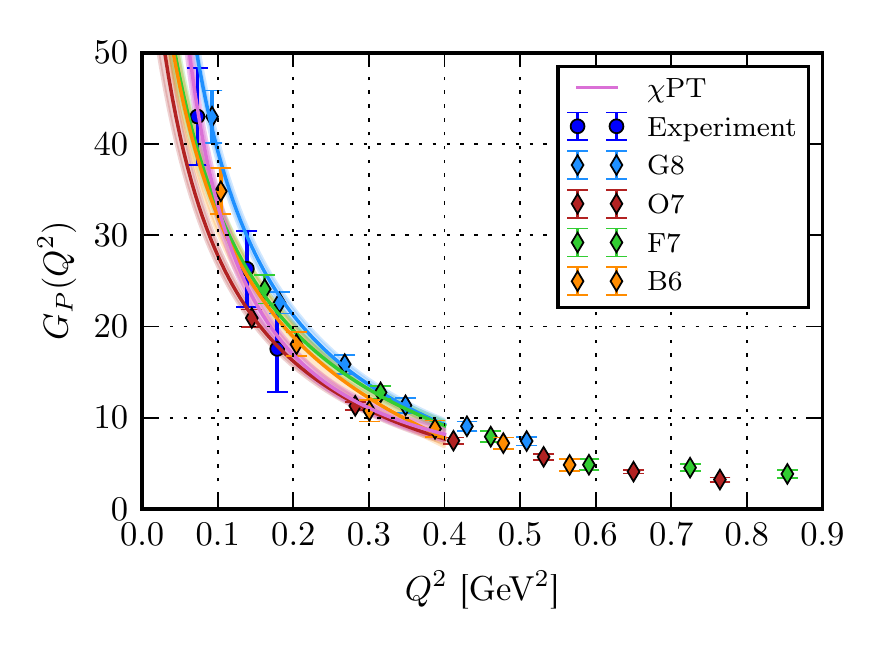}
\vspace{-0.3cm}
\end{center}
\caption{Our results for the induced pseudoscalar form factor $\GX{P}(Q^2)$ of the
nucleon, with the chiral fit and its extrapolation to the physical point.
The band in the same colour as the data point for each ensemble represents the result of
inserting into the fit function the pion mass and lattice spacing of that ensemble,
and the purple band labelled ``$\chi$PT'' represents the fit function in the continuum
limit at the physical pion mass.
Also shown (blue circles) are experimental results \cite{Choi:1993vt}, to which our
prediction can be seen to compare quite favourably.}
\label{fig:GPbigfit}
\end{figure}

Examples of the standard fit are shown in figures~\ref{fig:GAbigfit}
and \ref{fig:GPbigfit}, and the values for $g_{\rm A}$, $\langle
r_{\rm A}^2\rangle$, and $g_{\rm P}$ resulting at the physical pion
mass for the different variations are tabulated in
Table~\ref{tab:bigfit}.  The shape and overall normalization of
$\GX{P}$ in particular compares very favourably with the (limited)
experimental data. The axial charge in the chiral limit tends to come
out at a large value, while the obtained value of the parameter
$d_{16}$ gives an unnaturally large negative slope of $g_{\rm A}$ as a
function of $m_\pi^2$.  For this and other reasons given in section
\ref{sec:discussion}, we use fits to the pion-mass dependence of the
axial charge and radius for our final results; these fits are
presented in the next subsection.  Nonetheless, we think that the
effectiveness and stability of the EFT fit should be re-evaluated once
more accurate data at small pion mass and virtualities becomes
available.

\begin{table}
\caption
{Results for the axial charge $g_{\rm A}$, square of the axial charge radius
$\langle r_{\rm A}^2\rangle$, and pseudoscalar charge $g_{\rm P}$ of the nucleon from
a chiral EFT fit to our lattice data. The $\chi^2$ per degree of freedom and the number 
$\nu_{\rm d.o.f.}$ of degree of freedom are given in the last two columns. 
In all fits only the four most chiral ensembles G8, F7, F6 and B6 are used.
 The ``Standard'' fit incorporates a momentum cut $Q^2<0.4$~GeV$^2$, uses the lattice nucleon mass on each ensemble,
explicitly accounts for $\mathcal{O}(a^2)$ cut-off effects and has 9 fit parameters altogether. Also shown are
several variations that can be used to estimate the systematic
error.\vspace{16pt}}
{
\label{tab:bigfit}
\renewcommand\arraystretch{1.5}
\begin{tabular}{lcccccc}\hline\hline
Fit variant & Method & $g_{\rm A}$        & $\langle r_{\rm A}^2\rangle$ [fm$^2$] & $g_{\rm P}$ 
 & $\chi_{\rm red}^2$  &  $\nu_{\rm d.o.f.}$ \\\hline
Standard & summation & 1.255(71) & 0.238(77) & 8.3(5) & 0.59 & 19 \\
No $\mathcal{O}(a^2)$ & & 1.318(41) & 0.216(58) & 8.8(3) & 0.63 & 23\\
No $Q^2$ cut & & 1.325(48) & 0.217(29) & 8.8(3) & 0.61  & 43 \\
With $m_N^{\rm phys}$ & & 1.314(78) & 0.217(75) & 8.9(5) & 1.71 & 19\\
Alternative $c_i$ & & 1.336(71) & 0.209(73) & 8.8(5) & 0.47  & 19  \\
\hline
Standard & two-state & 1.382(91) & 0.246(92) & 9.0(7) & 0.50 & 19 \\
No $\mathcal{O}(a^2)$ & & 1.418(66) & 0.202(69) & 9.4(4) & 0.61  & 23 \\
No $Q^2$ cut & & 1.427(70) & 0.287(36) & 9.4(5) & 0.50 & 43 \\
With $m_N^{\rm phys}$ & & 1.399(98) & 0.214(92) & 9.4(7) & 1.25 & 19 \\
Alternative $c_i$ & & 1.478(91) & 0.222(86) & 9.7(7) & 0.29 & 19 \\
\hline\hline
\end{tabular}
}
\end{table}

\subsection{HBChPT-inspired fits}

In order to enable a comparison with the standard approach, we
also perform fits to the pion-mass dependence of the axial radius and the
axial charge using several variants of HBChPT-inspired chiral fits.
In order to determine whether our data allow for the resolution of the chiral
logarithm, we fit each quantity $\mathcal{Q}$ using the ansätze
\begin{align}
\label{eq:genfitlin}
\mathcal{Q}(m_\pi) &= A + B m_\pi^2,\\
\mathcal{Q}(m_\pi) &= A + B m_\pi^2 + C m_\pi^2\log m_\pi,
\label{eq:genfitlog}
\end{align}
with three and two fit parameters, respectively.
The linear fit (\ref{eq:genfitlin}) is applied both with (Fit~A) and without (Fit~B)
a pion-mass cut of $m_\pi\le 335$~MeV\footnote{The ensemble N6 is included both in Fit~A and in Fit~B.}  
in order to check for the importance of
higher-order corrections, while the fit (\ref{eq:genfitlog}) including a
logarithmic term is applied over the whole pion mass range (Fit~C).

In the case of the axial charge, we use a modified version of Fit~C, namely
\begin{equation}
g_{\rm A}(m_\pi) = \mathring{g}_{\rm A} + B m_\pi^2 - \frac{\mathring{g}_{\rm A}}{8\pi^2 F^2}\left(1+2\mathring{g}_{\rm A}^2\right)m_\pi^2\log m_\pi
\label{eq:xptgafit}
\end{equation}
with fit parameters $\mathring{g}_{\rm A}$ and $B$, and the chiral-limit pion decay
constant fixed to its phenomenological value
$F=86$~MeV~\cite{Colangelo:2003hf,Aoki:2016frl};
a pion-mass cut of $m_\pi\le 335$~MeV is applied in this case.
The reason is that using Eq.~(\ref{eq:genfitlog}) with the coefficient of the
logarithmic term left free gives implausible results in that the sign
of the logarithmic term comes out positive, contrary to the ChPT result
incorporated into Eq.~(\ref{eq:xptgafit}).
Comparing the results of the different fits (cf. Fig.~\ref{fig:gAextrapolation}),
we conclude that our data are not precise enough to allow for a reliable
resolution of the chiral logarithm, and that applying the fit form  \ref{eq:xptgafit}
amounts to imposing a trend which is not seen at all in our data.
Even with Fit A, we observe that the slope of $g_A$ as a function of $m_\pi^2$ 
is only poorly constrained. 
In the results presented in Table~\ref{tab:chiralfits}, 
we have assumed that the results are independent of the lattice spacing,
since we already observed that no particular trend in the cutoff dependence 
is seen in the $g_A$ or $\langle r_A^2\rangle$ data for $m_\pi\leq 335\,$MeV.
For an indication of how sensitive the results are to this assumption,
we quote the result of a simultaneous linear extrapolation in $a^2$ and $m_\pi^2$ 
(i.e.\ a three-parameter fit) of the axial charge 
for $m_\pi\leq 335\,$MeV: we find $g_A = 1.236(88)$ at the physical pion mass,
to be compared with $g_A = 1.278(68)$ assuming no cutoff effects. 
The difference is well contained in the uncertainty estimate.
At the same time, we note that the description including O($a^2$) cutoff effects
is overfitting the data, as witnessed by the fact that the $\chi^2/{\rm d.o.f.}$ goes up
from 1.07 to 1.21 in spite of the additional fit parameter. 
As a further alternative, if we perform Fit A without an O($a^2)$ but remove 
the data from the coarsest lattice spacing, we obtain $g_A=1.278(72)$ at the physical pion mass,
i.e.\ no change except for an insignificant increase in the statistical uncertainty.

\begin{table}
\caption{Summary of the results of chiral fits using a linear fit form
 with a pion-mass cut of $m_\pi\le 335$~MeV (Fit A) or no pion-mass cut
 (Fit B), or a ChPT-inspired fit form with a logarithmic term (Fit C).
 In the case of $g_{\rm A}$, Fit C uses Eq.~(\ref{eq:xptgafit}) with a pion mass
 cut of $m_\pi\le 335$~MeV; otherwise, Fit C uses Eq.~(\ref{eq:genfitlog}) and
 data at all pion masses.\vspace{16pt}}
{
\label{tab:chiralfits}
\renewcommand\arraystretch{1.5}
\begin{tabular}{lccc}
\hline\hline
      & $g_{\rm A}$     & $\left\langle r_{\rm A}^2\right\rangle$ [fm$^2$] & $g_{\rm P}$  \\\hline
      & \multicolumn{3}{c}{Two-state fit}\\\hline
Fit A~~~~~ & 1.278(68) & 0.360(36)                         & 7.7(1.8)   \\
Fit B & 1.191(27) & 0.271(12)                         & 8.5(1.5)   \\
Fit C & 1.186(56) & 0.440(47)                         & 5.7(2.1)   \\\hline
      & \multicolumn{3}{c}{Summation method}\\\hline
Fit A & 1.208(69) & 0.279(26)                         & 8.2(1.6)   \\
Fit B & 1.200(34) & 0.242(12)                         & 8.0(1.5)   \\
Fit C & 1.138(59) & 0.330(39)                         & 7.6(1.9)  \\\hline\hline
\end{tabular}
}
\end{table}

For the axial charge radius, we find that the linear fits without a
pion-mass cut do not describe our
data well, whereas both the linear fits with a pion-mass cut and the
fits including a logarithmic term are well compatible with our data
(cf. Fig.~\ref{Fig:rad_zexp}).

\begin{figure}
\begin{center}
\includegraphics[width=0.5\textwidth,keepaspectratio=]{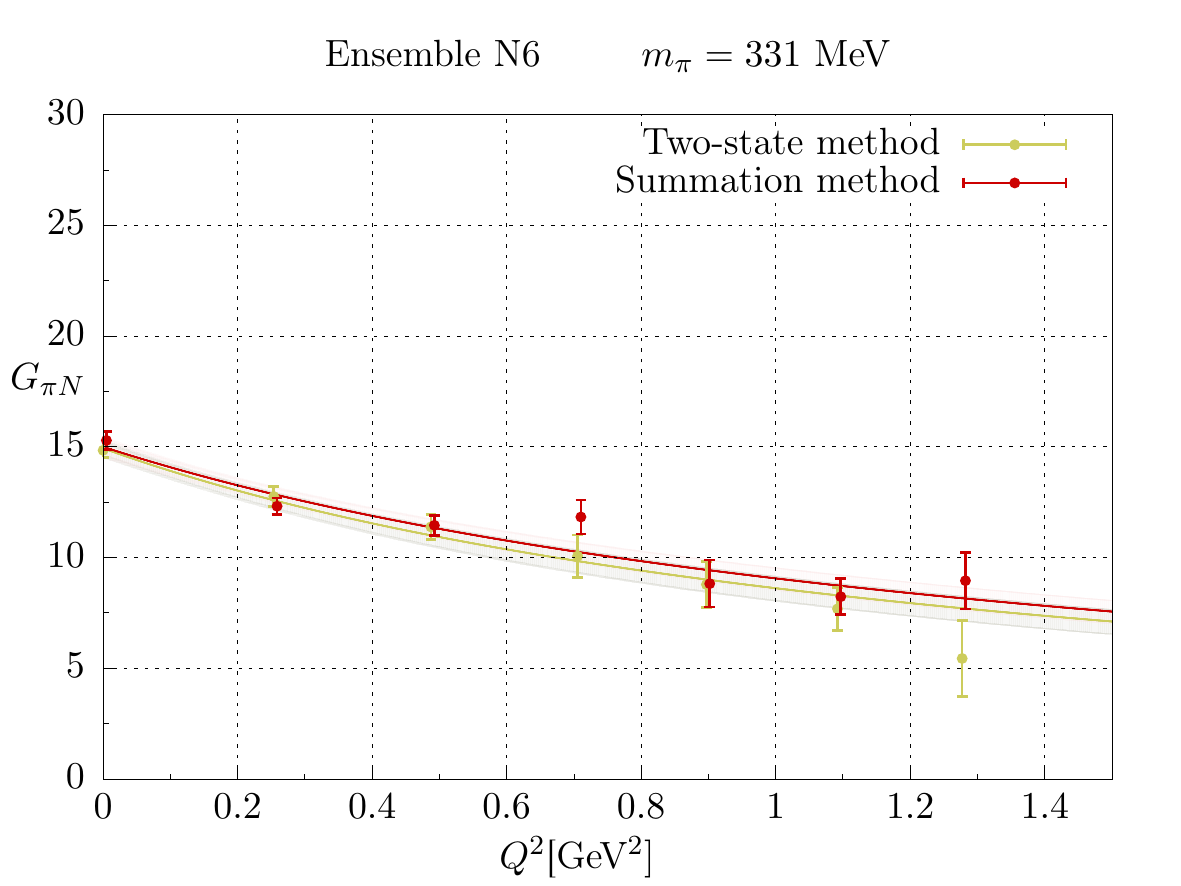}
\end{center}
\vspace{-0.6cm}
\caption{The form factor $G_{\pi N}(Q^2)$ on ensemble N6, parameterized by the monopole ansatz
(\ref{eq:gpiNansatz}), for the summation and the two-state-fit methods.}
\label{fig:GpiN_N6}
\end{figure}

\begin{figure}
\begin{center}
\includegraphics[width=0.48\textwidth,keepaspectratio=]{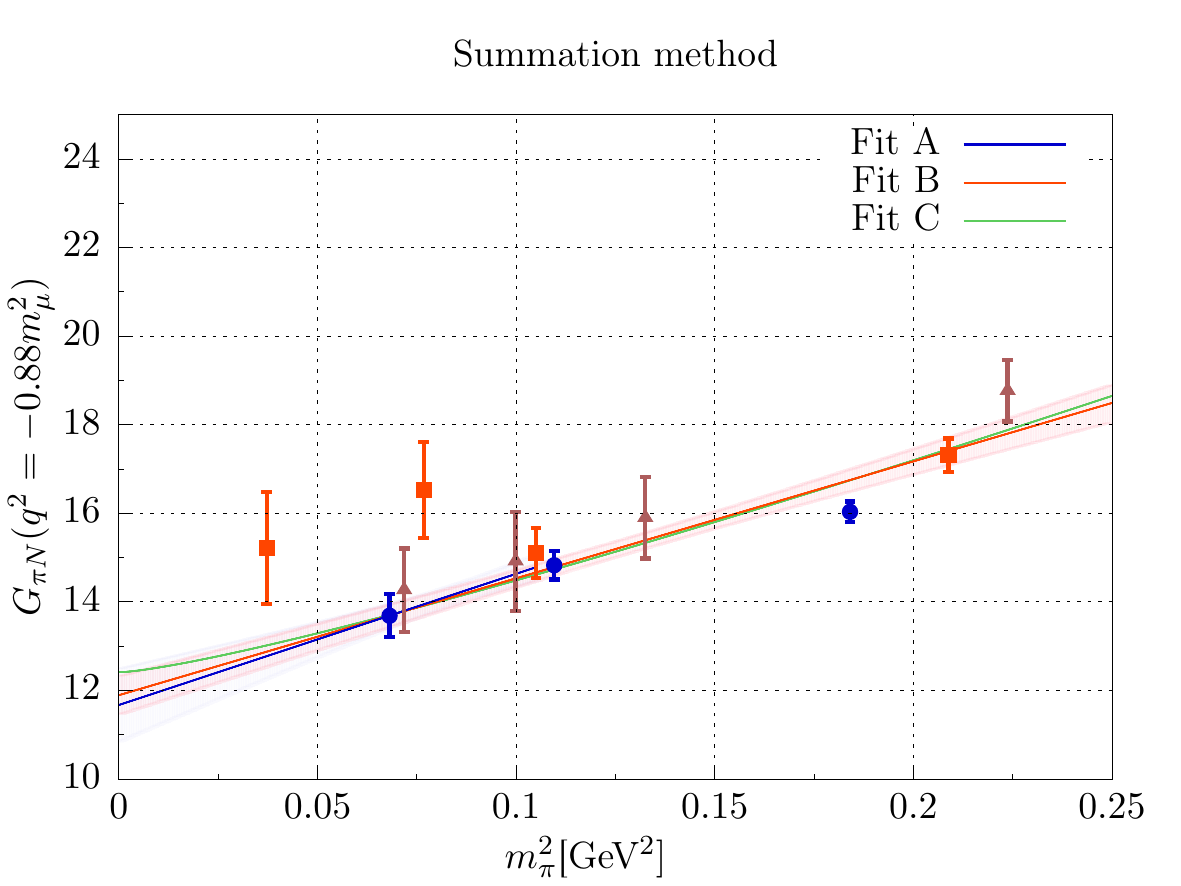}
\includegraphics[width=0.48\textwidth,keepaspectratio=]{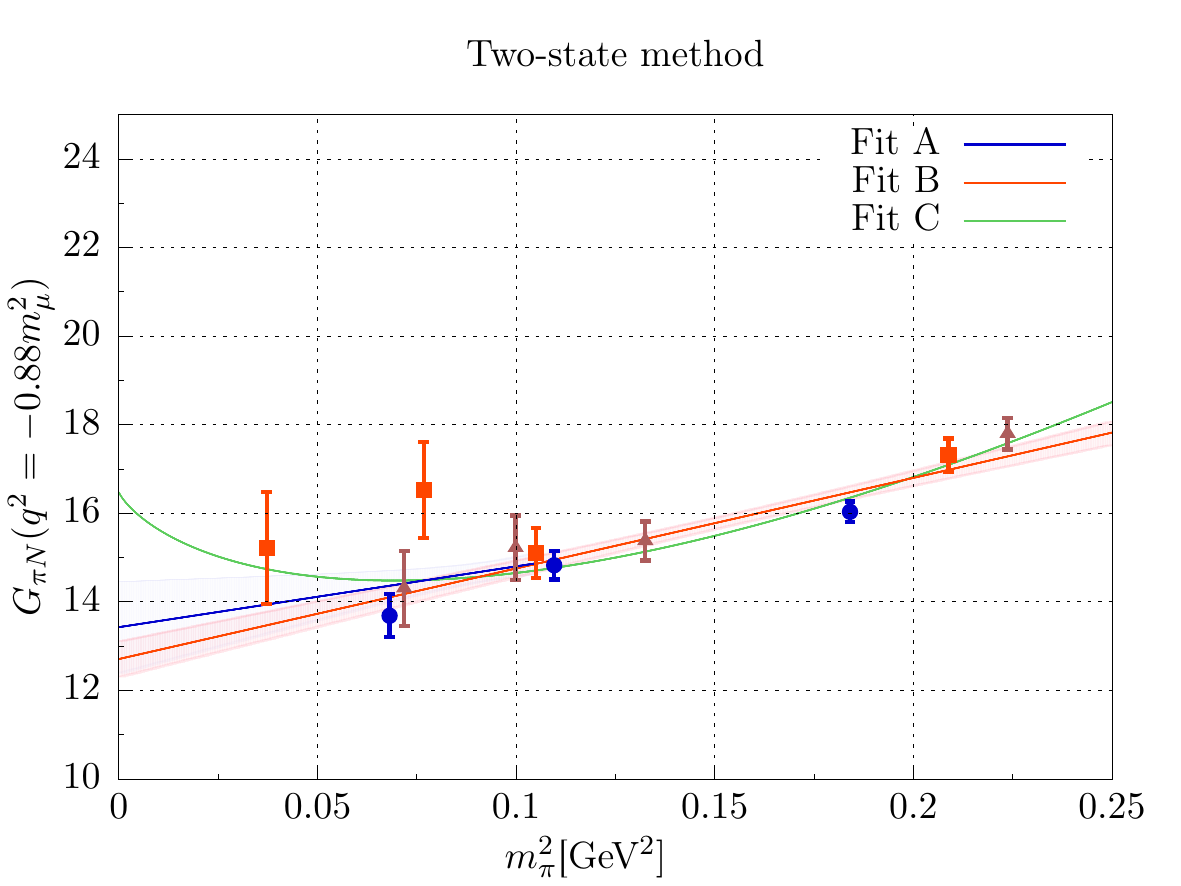}
\end{center}
\vspace{-0.6cm}
\caption{Chiral extrapolation of $G_{\pi N}(Q_*^2)$
         for the summation method (left) and 
         the two-state-fit method (right panel).
 Triangles, squares and circles correspond to increasingly fine lattice spacings.
 Fit A is a linear fit with a pion-mass cut $m_\pi\leq 335{\rm\,MeV}$,
 Fit B is a linear fit with no pion-mass cut, and
 Fit C incorporates a logarithmic term via eqn.~(\ref{eq:genfitlog}).
}
\label{fig:GpiNextrap}
\end{figure}

The pseudoscalar coupling defined by Eq.~(\ref{Eq:smallgp})
is not readily accessible from our results for $\GX{P}$ because 
its $Q^2$ and $m_\pi^2$ dependence is strong, due to the pion pole.
However, observing that the pion-nucleon form factor $G_{\pi N}(Q^2)$ 
depends less strongly on $Q^2$ and $m_\pi^2$, we proceed as follows.
First, the form factor $G_{\pi N}(Q^2)$ is determined on every ensemble 
at the available $Q^2$ values by taking the appropriate linear combination of 
$\GX{A}(Q^2)$  and $\GX{P}(Q^2)$; see Eqs.\ (\ref{Eq:FP_GpiN}) and (\ref{Eq:pcac_mat}).
Then a monopole fit is performed (see Fig. \ref{fig:GpiN_N6}), 
\begin{equation}
G_{\pi N}(Q^2) = \frac{C}{\Lambda^2+Q^2},
\label{eq:gpiNansatz}
\end{equation}
which allows us to extract $G_{\pi N}(Q_*^2)$ on every ensemble.
The latter is then chirally (and continuum) extrapolated to the physical point; see Fig. \ref{fig:GpiNextrap}.
We use the already determined values of $g_{\rm A}$ and $\left\langle r_{\rm A}^2\right\rangle$
at the physical pion mass to obtain $\GX{A}(Q_*^2)$ via (\ref{eq:GAtaylor}), since $Q_*^2$ is very small.
Finally, taking the appropriate linear combination of  $\GX{A}(Q_*^2)$ and $G_{\pi N}(Q_*^2)$
yields $g_{\rm P}$ at the physical pion mass.
The momentum-transfer dependence (\ref{eq:gpiNansatz}) of $G_{\pi N}$ was proposed in~\cite{Schindler:2006it};
it describes the lattice data well, as can be seen from Fig. \ref{fig:GpiN_N6}.
We find that this dependence is very mild,
reflecting the fact that the pion pole describes the bulk of the $Q^2$ dependence of $F_{\rm P}(Q^2)$. 
Since $Q_*^2$ is small on hadronic scales and  Eq.\ \ref{Eq:pcac_mat} implies $G_{\pi N}(0) = \frac{M_N g_{\rm A}}{F_\pi}$,
the sensitivity of $G_{\pi N}(Q_*^2)$ to the specific form of the ansatz (\ref{eq:gpiNansatz}) is weak.

A summary of our results for $g_{\rm A}$, $\left\langle r_{\rm A}^2\right\rangle$,
and $g_{\rm P}$ can be found in Table~\ref{tab:chiralfits}.


\section{Discussion and Conclusion}
\label{sec:discussion}

As discussed above, we use in our main analysis the form factors
extracted with the two-state fit in order to remove excited-state contributions.
We include the pseudoscalar density in our analysis, as its matrix elements 
are overall compatible with those of the axial current via the PCAC relation
and tend to increase the precision of the calculation.

As for the chiral extrapolation, we have performed simultaneous fits
to the pion-mass and momentum transfer dependence, as well as the more
widely used two-step procedure of first extracting the small-$Q^2$
observables and then chirally extrapolating them.
A disadvantage of the simultaneous chiral fits to $\GX{A}$ and $\GX{P}$ is that it is
intrinsically a small-$Q^2$ expansion, and the paucity
of the lattice data in this region, as compared to previous lattice
calculations of the pion electromagnetic form
factor~\cite{Brandt:2013dua}, adversely affects the stability of the
fits. Empirically, we find that the fitted values of the low-energy constants
$d_{16}$ and $d_{18}$ come out large and poorly constrained, casting some doubt
on whether convergence is under control. 
Secondly, in the chiral expansion the
leading correction to the (pion-mass independent) leading-order value
for the slope of the axial form factor happens to vanish; this implies
that the fit ansatz imposes a pion-mass dependence of the axial radius
which is completely dominated by the pion-mass dependence of the axial
charge (see Eq.\ (\ref{Eq:radius})).
Third, the chiral logarithm predicted in the pion-mass dependence of the axial
charge is not seen in the data (see Fig.\ \ref{fig:gAextrapolation}).
For these reasons, unlike in our study of the vector form
factors~\cite{Capitani:2015sba}, we prefer in the present study to
quote as final results those obtained with the more conventional
procedure of first extracting the axial charge and radius and $g_{\rm P}$ on
each ensemble, followed by a chiral extrapolation.
Nonetheless, we have presented the effective field theory analysis because this type of combined fits
may be useful in future calculations involving more accurate data in the very chiral and low-momentum regime.

For our final numbers, we choose to quote the result of applying a linear fit
with a pion-mass cut of $m_\pi\le 335$\,MeV (Fit A) to the data obtained from
the two-state fit method. We estimate the systematic error from the chiral extrapolation
by taking the difference between the fits with and without a pion-mass cut (Fits A and B)
as a one-sided systematic error; where the fit including a logarithmic term (Fit C)
lies on the other side of our central value, we include the corresponding difference
into an asymmetric two-sided systematic error. We note that the results from the
summation method are also covered by the resulting error bars, and that our results
are therefore not sensitive to excited-state effects at this level of accuracy.
We thus finally obtain
\begin{eqnarray}
g_{\rm A} &=& 1.278 \pm 0.068\genfrac{}{}{0pt}{1}{+0.000}{-0.087},\nonumber
\\
\langle r_{\rm A}^2\rangle &=& 0.360 \pm 0.036\genfrac{}{}{0pt}{1}{+0.080}{-0.088}~\mathrm{fm}^2,
\\
g_{\rm P} &=& 7.7 \pm 1.8 \genfrac{}{}{0pt}{1}{+0.8}{-2.0} ,\nonumber
\end{eqnarray}
where the first error is statistical and the second systematic.
All three results are compatible with the phenomenological values of these quantities:
$g_A = 1.2723(23)$ (the Particle Data Group average of neutron $\beta$ decay data \cite{Olive:2016xmw}), 
$\langle r_{\rm A}^2\rangle = 0.444(18){\rm fm}^2$ (from a dipole fit to neutrino scattering data, see \cite{Bernard:2001rs} 
for a review and other determinations of the axial charge radius),
and $g_{\rm P}= 8.06(55)$ (from the MuCap experiment~\cite{Andreev:2012fj,Andreev:2015evt}).

Several other lattice calculations of the axial charge have appeared
recently \cite{Bali:2014nma,Abdel-Rehim:2015owa,Bhattacharya:2016zcn,
Yoon:2016jzj,Berkowitz:2017gql,Meyer:2016kwb,Alexandrou:2017msl,
Green:2017keo,Alexandrou:2017hac,Rajan:2017lxk}.
In particular, the values obtained in~\cite{Bali:2014nma,Abdel-Rehim:2015owa,
Bhattacharya:2016zcn,Yoon:2016jzj,Berkowitz:2017gql}
at the physical pion mass also agree with the phenomenological value,
while the calculation in \cite{Alexandrou:2017hac} yields a slightly lower value.
Most of them quote a
rather more precise result, however the precision depends strongly on
the source-sink separations and the ranges of pion masses used in the
chiral extrapolation.  For example, as compared to our earlier
publication on the axial charge~\cite{Capitani:2012gj}, the final
quoted error has changed little, but the present chiral extrapolation
is based on the interval $190\leq m_\pi/{\rm MeV}\leq 335$, rather than on $270\leq
m_\pi/{\rm MeV}\leq 540$. This comparison also illustrates the 
increasing computational cost of determining nucleon structure observables
as the pion mass is reduced. 

\begin{figure}
\begin{center}
\includegraphics[width=0.5\textwidth,keepaspectratio=]{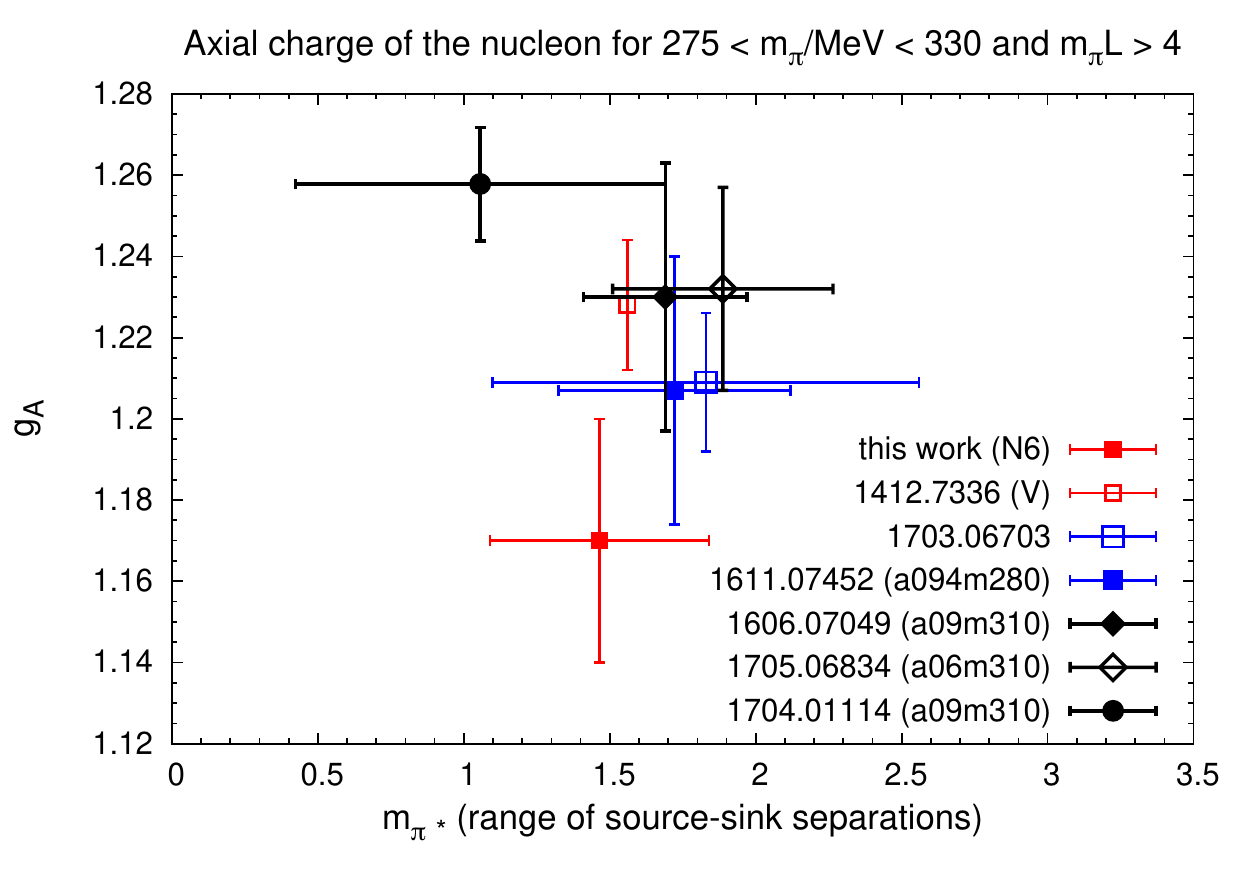}
\end{center}
\vspace{-0.4cm}
\caption{Comparison of select recent lattice results for the axial charge of the nucleon at $m_\pi\simeq 300\,$MeV.
Attached to each data point, a horizontal line indicates the range of source-sink separations used in the calculation.
The caption indicates the arXiv number of the respective publications, with the relevant lattice ensemble given in brackets.
Red points correspond to two-flavor, blue points to 2+1 flavor and black points to 2+1+1 flavor calculations.
Square data points correspond to an O($a$) improved Wilson fermion action; the other data points come from a mixed-action
approach with staggered sea quarks.
}
\label{fig:compa}
\end{figure}

In order to remove the effect of applying different
procedures for the chiral extrapolations, in Fig.\ \ref{fig:compa} we compare 
recent lattice results for the axial charge at $m_\pi\simeq300\,$MeV obtained 
on reasonably large and fine lattices satisfying $m_\pi L>4$ and $a<0.095\,$fm.
Since the range of source-sink separations used in the calculations is one of the main 
quality criteria, we indicate this range by a horizontal segment in Fig.\ \ref{fig:compa}.
Within the uncertainties of the comparison, we observe a consistent picture. 
Most results cluster between 1.20 and 1.24, while the central value of 
our result on ensemble N6 lies somewhat lower. 
While the available data does not allow one to identify a specific trend as a function of the 
flavor content of the calculations or on the lattice spacing,
future continuum-extrapolated results would allow for a more controlled comparison.
Finally, we remark that the challenge for the future will be to maintain or improve the statistical precision
while extracting the axial charge exclusively from source-sink separations exceeding 1.5\,fm.

To a more limited extent, we can also compare our result for the axial
radius to other lattice determinations. Again, we favor performing the
comparison at a fixed pion mass of about 300\,MeV. In extracting the
radius, the precise fit form used to describe the axial form factor
plays an important role, both for the central value and the achieved
statistical precision. Using a dipole fit, Ref.\ \cite{Rajan:2017lxk}
obtains $\langle r_A^2\rangle = 0.242(29)\,{\rm fm}^2$ for
$m_\pi=320\,{\rm MeV}$ on their ensemble a06m310; with a dipole fit,
our result $\langle r_A^2\rangle = 0.239(21)\,{\rm fm}^2$ for
$m_\pi=331\,{\rm MeV}$ on ensemble N6 agrees very well. Also, at
$m_\pi=317\,{\rm MeV}$ Ref.\ \cite{Green:2017keo} obtains $0.213(15)$,
where we have added the errors in quadrature, from a $z$-expansion fit
with five terms and Gaussian priors starting at the third term.  These
recent results are thus in good agreement.

Probably the most critical step in the presented lattice calculation
is the extraction of the ground-state matrix elements from the
correlation functions.  We have used two theoretically well motivated
methods, the summation and two-state method described in section
\ref{sec:excited}, in order to perform this task. For the axial form
factor at $Q^2\lesssim0.5{\rm GeV}^2$, we observe good agreement
between the two methods, adding confidence in the results for $g_A$
and $\langle r_A^2\rangle$.  The situation for the form factors
containing the pion pole ($\GX{P}$, $F_{\rm P}$) is less satisfactory,
in the sense that only the two-state-fit method led to ground-state
matrix elements consistent with the PCAC relation
Eq.\ (\ref{Eq:pcac_mat}). It is for that reason that we chose the
two-state-fit method to present our final results. In the future, by
performing calculations at larger source-sink separations, we hope to
benefit from more cross-checks, as in the case of the axial form
factor. At the same time, the corrections to the mid-point values of
the ratios $R_.(t=t_s/2,t_s)$, induced by the summation or
two-state-fit method in extracting the ground-state matrix elements,
would be reduced.

In summary, we have performed a two-flavour lattice QCD calculation of
the isovector axial and induced pseudoscalar nucleon form factors. We
have consistently applied the $\mathcal{O}(a)$ improvement program and observe
no significant cutoff effect on the low-$Q^2$ observables at pion
masses below 350\,MeV.  We have made use of the pseudoscalar
density. Its form factor is related to the axial and induced
pseudoscalar form factor by the PCAC relation, and thus it provides
both a cross-check of the calculation and a slight increase in
precision. The axial charge, the axial radius and the pseudoscalar
coupling we obtained are in agreement with experiment.

In the near future, we plan to improve on our calculation by using
$2+1$-flavour ensembles (i.e.\ containing the sea-quark effects of
both the light and strange quarks), by increasing the statistics by an
order of magnitude and by going to larger source-sink separations. A
preliminary account of the results was presented at the Lattice 2016
conference~\cite{Djukanovic:2016ocj}.  The lessons learnt from the
analysis applied here will be beneficial in this endeavour.

\section*{Acknowledgments}

We are grateful to our colleagues within the CLS initiative for
sharing ensembles.
We have used the LoopTools library \cite{Hahn:1998yk} in the numerical evaluation of
the EFT amplitudes \cite{Schindler:2006it}.

Our calculations were performed on the ``Wilson''
HPC Cluster at the Institute for Nuclear Physics, University of Mainz,
and on the ``Clover'' HPC Cluster at the Helmholtz Institute Mainz.
We are grateful to Christian Seiwerth for technical support.
We are also grateful for
computer time allocated to project HMZ21 on the BG/Q “JUQUEEN”
computer at NIC, J\"ulich. This work was granted access to the HPC
resources of the Gauss Center for Supercomputing at Forschungzentrum
J\"ulich, Germany, made available within the Distributed European
Computing Initiative by the PRACE-2IP, receiving funding from the
European Community’s Seventh Framework Programme (FP7/2007-2013) under
Grant No.\ RI-283493.

This work was supported by the Deutsche Forschungsgemeinschaft (DFG) in the
SFB\,443 and SFB\,1044,
and by the Rhineland-Palatinate Research Initiative.
M.D.M.\ was partially supported by the Danish National Research Foundation
under Grant No.\ DNRF:90.
B.J.\ was supported by the Schweizerischer Nationalfonds (SNF) under grant 200020-162515.
P.M.J.\ acknowledges support from the Department of Theoretical Physics, TIFR.
T.D.R.\ was supported by DFG grant HA4470/3-1.

\appendix

\section{Form factor values}
\label{app:tables}

In Tables~%
\ref{tab:A3}--\ref{tab:O7}
we give all of our results for the iso-vector axial form factors $\GX{A}$
and $\GX{P}$ of the nucleon at all values of $Q^2$ measured on each ensemble.
Listed in each case are the values obtained using the summation method and an
explicit two-state fit (cf. the main text for details).
The statistical errors on each data point are quoted in parentheses following
the central value.

\renewcommand\arraystretch{1.2}

\begin{table*}[h!]\small
\caption
{
A3 ensemble ($a=0.079\,$fm, $m_{\pi}=473\,$MeV, $t_s/a\in\{10,12,14,16\}$): The axial, induced pseudoscalar, and pion-nucleon pseudoscalar form factors at all $Q^2$ values for all extraction methods.}
{
\renewcommand{\multirowsetup}{\centering}
\begin{tabular}{|c|c|c|c|c|c|c|}
        \hline
        A3 & \multicolumn{2}{c|}{$G_{\rm A}$} & \multicolumn{2}{c|}{$G_{\rm P}$} & \multicolumn{2}{c|}{$G_{\pi N}$} \\
        \hline
        $Q^{2}$[$\text{GeV}^2$] & Two-state & Summation & Two-state & Summation & Two-state & Summation \\
        \hline
        0.0   & 1.222 (0.024) & 1.276 (0.049) &      -        &     -         & 18.03 (0.356) & 18.83 (0.717) \\
        0.230 & 1.057 (0.025) & 1.052 (0.041) & 17.47 (0.664) & 16.88 (1.125) & 15.43 (0.413) & 15.84 (0.637) \\
        0.448 & 0.922 (0.026) & 0.919 (0.040) & 9.829 (0.345) & 10.14 (0.568) & 14.63 (0.598) & 13.68 (0.805) \\
        0.655 & 0.790 (0.033) & 0.755 (0.054) & 6.222 (0.338) & 6.034 (0.575) & 14.04 (0.833) & 12.99 (1.313) \\
        0.823 & 0.680 (0.043) & 0.597 (0.074) & 4.656 (0.381) & 4.057 (0.678) & 11.37 (1.011) & 10.27 (1.812) \\
        1.013 & 0.675 (0.051) & 0.612 (0.084) & 3.955 (0.323) & 3.722 (0.561) & 11.13 (1.080) & 8.616 (1.793) \\
        1.196 & 0.585 (0.067) & 0.494 (0.115) & 2.858 (0.357) & 2.466 (0.649) & 11.75 (1.694) & 9.133 (2.869) \\
        \hline
\end{tabular}
\label{tab:A3}
}
\end{table*}

\begin{table*}[h!]\small
\caption{
A4 ensemble ($a=0.079\,$fm, $m_{\pi}=364\,$MeV, $t_s/a\in\{10,12,14,16\}$): The axial, induced pseudoscalar, and pion-nucleon pseudoscalar form factors at all $Q^2$ values for all extraction methods.}
{
\renewcommand{\multirowsetup}{\centering}
\begin{tabular}{|c|c|c|c|c|c|c|}
        \hline
        A4 & \multicolumn{2}{c|}{$G_{\rm A}$} & \multicolumn{2}{c|}{$G_{\rm P}$} & \multicolumn{2}{c|}{$G_{\pi N}$} \\
        \hline
        $Q^{2}$[$\text{GeV}^2$] & Two-state & Summation & Two-state & Summation & Two-state & Summation \\
        \hline
        0.0   & 1.170 (0.033) & 1.199 (0.080) &     -         &       -       & 15.42 (0.430) & 15.80 (1.054) \\
        0.229 & 1.022 (0.038) & 0.985 (0.042) & 16.41 (1.069) & 15.74 (1.135) & 14.04 (0.826) & 13.64 (0.914) \\
        0.442 & 0.978 (0.054) & 0.984 (0.074) & 9.304 (0.541) & 9.258 (0.771) & 16.35 (2.337) & 16.90 (2.163) \\
        0.643 & 0.823 (0.056) & 0.817 (0.074) & 6.064 (0.492) & 5.812 (0.670) & 12.92 (1.242) & 14.51 (2.047) \\
        0.805 & 0.926 (0.164) & 0.822 (0.161) & 6.017 (1.184) & 5.299 (1.133) & 10.39 (1.894) & 9.725 (2.688) \\
        0.987 & 0.674 (0.088) & 0.546 (0.103) & 3.429 (0.486) & 2.857 (0.584) & 11.63 (1.648) & 7.992 (2.287) \\
        1.162 & 0.575 (0.106) & 0.359 (0.159) & 2.537 (0.509) & 1.830 (0.712) & 10.18 (1.893) & 0.205 (4.267) \\
        \hline
\end{tabular}
\label{tab:A4}
}
\end{table*}

\begin{table*}[h!]\small
\caption{A5 ensemble ($a=0.079\,$fm, $m_{\pi}=316\,$MeV, $t_s/a\in\{10,12,14,16\}$): The axial, induced pseudoscalar, and pion-nucleon pseudoscalar form factors at all $Q^2$ values for all extraction methods.}
{
\renewcommand{\multirowsetup}{\centering}
\begin{tabular}{|c|c|c|c|c|c|c|}
        \hline
        A5 & \multicolumn{2}{c|}{$G_{\rm A}$} & \multicolumn{2}{c|}{$G_{\rm P}$} & \multicolumn{2}{c|}{$G_{\pi N}$} \\
        \hline
        $Q^{2}$[$\text{GeV}^2$] & Two-state & Summation & Two-state & Summation & Two-state & Summation \\
        \hline
        0.0   & 1.208 (0.058) & 1.238 (0.092) &      -        &       -       & 15.27 (0.738) & 15.65 (1.165) \\
        0.228 & 1.018 (0.069) & 0.987 (0.064) & 16.96 (1.541) & 16.90 (1.462) & 12.88 (0.941) & 11.72 (1.003) \\
        0.440 & 0.853 (0.072) & 0.869 (0.078) & 8.433 (0.822) & 8.662 (0.934) & 11.94 (1.111) & 11.74 (1.056) \\
        0.638 & 0.620 (0.088) & 0.642 (0.099) & 4.386 (0.739) & 4.245 (0.816) & 10.13 (1.910) & 13.71 (2.561) \\
        0.797 & 0.451 (0.143) & 0.424 (0.162) & 2.343 (0.946) & 2.067 (1.105) & 12.45 (2.555) & 13.94 (3.250) \\
        0.977 & 0.410 (0.094) & 0.357 (0.115) & 2.075 (0.485) & 1.775 (0.616) & 5.331 (2.239) & 5.396 (2.410) \\
        1.148 & 0.389 (0.149) & 0.376 (0.187) & 1.731 (0.641) & 1.451 (0.813) & 4.088 (5.517) & 11.20 (5.863) \\
        \hline
\end{tabular}
\label{tab:A5}
}
\end{table*}

\begin{table*}[h!]\small
\caption{B6 ensemble ($a=0.079\,$fm, $m_{\pi}=268\,$MeV, $t_s/a\in\{10,12,14,16\}$): The axial, induced pseudoscalar, and pion-nucleon pseudoscalar form factors at all $Q^2$ values for all extraction methods.}
{
\renewcommand{\multirowsetup}{\centering}
\begin{tabular}{|c|c|c|c|c|c|c|}
        \hline
        B6 & \multicolumn{2}{c|}{$G_{\rm A}$} & \multicolumn{2}{c|}{$G_{\rm P}$} & \multicolumn{2}{c|}{$G_{\pi N}$} \\
        \hline
        $Q^{2}$[$\text{GeV}^2$] & Two-state & Summation & Two-state & Summation & Two-state & Summation \\
        \hline
        0.0   & 1.214 (0.065) & 1.275 (0.080) &      -        &       -       & 14.56 (0.781) & 15.29 (0.957) \\
        0.104 & 1.158 (0.095) & 1.120 (0.074) & 35.15 (3.090) & 34.82 (2.519) & 12.16 (1.285) & 11.26 (0.986) \\
        0.204 & 1.053 (0.094) & 1.031 (0.069) & 18.38 (1.807) & 18.06 (1.322) & 13.36 (1.441) & 12.99 (1.084) \\
        0.301 & 0.894 (0.112) & 0.886 (0.082) & 11.44 (1.549) & 10.78 (1.173) & 12.14 (2.959) & 14.13 (1.857) \\
        0.387 & 0.830 (0.096) & 0.794 (0.072) & 9.049 (1.142) & 8.792 (0.898) & 9.013 (1.865) & 7.812 (1.537) \\
        0.478 & 0.877 (0.092) & 0.812 (0.065) & 7.807 (0.833) & 7.222 (0.614) & 10.87 (2.163) & 10.16 (1.061) \\
        0.566 & 0.743 (0.119) & 0.679 (0.082) & 5.606 (0.843) & 4.837 (0.671) & 10.44 (5.042) & 13.06 (2.402) \\
        \hline
\end{tabular}
\label{tab:B6}
}
\end{table*}

\begin{table*}[h!]\small
\caption{E5 ensemble ($a=0.063\,$fm, $m_{\pi}=457\,$MeV, $t_s/a\in\{11,13,15,17\}$): The axial, induced pseudoscalar, and pion-nucleon pseudoscalar form factors at all $Q^2$ values for all extraction methods.}
{
\renewcommand{\multirowsetup}{\centering}
\begin{tabular}{|c|c|c|c|c|c|c|}
        \hline
        E5 & \multicolumn{2}{c|}{$G_{\rm A}$} & \multicolumn{2}{c|}{$G_{\rm P}$} & \multicolumn{2}{c|}{$G_{\pi N}$} \\
        \hline
        $Q^{2}$[$\text{GeV}^2$] & Two-state & Summation & Two-state & Summation & Two-state & Summation \\
        \hline
        0.0   & 1.174 (0.025) & 1.213 (0.047) &      -        &      -        & 17.54 (0.368) & 18.12 (0.695) \\
        0.357 & 0.958 (0.026) & 0.971 (0.038) & 12.85 (0.521) & 13.11 (0.752) & 14.47 (0.480) & 14.52 (0.686) \\
        0.685 & 0.741 (0.028) & 0.727 (0.040) & 5.680 (0.290) & 5.434 (0.419) & 14.76 (0.615) & 15.28 (0.849) \\
        0.991 & 0.621 (0.041) & 0.560 (0.065) & 3.644 (0.307) & 3.102 (0.486) & 12.67 (0.968) & 13.47 (1.596) \\
        1.236 & 0.575 (0.057) & 0.535 (0.099) & 2.816 (0.317) & 2.620 (0.566) & 12.27 (1.434) & 11.45 (2.540) \\
        1.511 & 0.491 (0.053) & 0.413 (0.090) & 2.064 (0.238) & 1.676 (0.411) & 10.08 (1.264) & 9.950 (2.194) \\
        1.772 & 0.455 (0.071) & 0.397 (0.130) & 1.627 (0.275) & 1.390 (0.504) & 10.93 (1.975) & 10.48 (3.452) \\
        \hline
\end{tabular}
\label{tab:E5}
}
\end{table*}

\begin{table*}[h!]\small
\caption{F6 ensemble ($a=0.063\,$fm, $m_{\pi}=324\,$MeV, $t_s/a\in\{11,13,15,17\}$): The axial, induced pseudoscalar, and pion-nucleon pseudoscalar form factors at all $Q^2$ values for all extraction methods.}
{
\renewcommand{\multirowsetup}{\centering}
\begin{tabular}{|c|c|c|c|c|c|c|}
        \hline
        F6 & \multicolumn{2}{c|}{$G_{\rm A}$} & \multicolumn{2}{c|}{$G_{\rm P}$} & \multicolumn{2}{c|}{$G_{\pi N}$} \\
        \hline
        $Q^{2}$[$\text{GeV}^2$] & Two-state & Summation & Two-state & Summation & Two-state & Summation \\
        \hline
        0.0   & 1.169 (0.041) & 1.165 (0.055) &       -       &      -        & 15.12 (0.529) & 15.07 (0.712) \\
        0.163 & 1.034 (0.059) & 1.014 (0.048) & 21.60 (1.655) & 21.69 (1.397) & 13.88 (0.667) & 13.15 (0.577) \\
        0.317 & 0.887 (0.052) & 0.886 (0.045) & 11.23 (0.754) & 11.02 (0.676) & 13.81 (0.842) & 14.37 (0.660) \\
        0.464 & 0.781 (0.062) & 0.801 (0.056) & 7.279 (0.694) & 6.993 (0.599) & 13.41 (1.128) & 16.45 (1.159) \\
        0.595 & 0.677 (0.069) & 0.690 (0.066) & 5.124 (0.663) & 5.297 (0.661) & 12.42 (1.016) & 12.02 (1.061) \\
        0.731 & 0.598 (0.058) & 0.653 (0.055) & 3.857 (0.423) & 4.217 (0.393) & 10.90 (0.953) & 11.79 (0.932) \\
        0.862 & 0.556 (0.078) & 0.601 (0.076) & 3.118 (0.494) & 3.331 (0.479) & 10.35 (1.070) & 11.94 (1.265) \\
        \hline
\end{tabular}
\label{tab:F6}
}
\end{table*}

\begin{table*}[h!]\small
\caption{F7 ensemble ($a=0.063\,$fm, $m_{\pi}=277\,$MeV, $t_s/a\in\{11,13,15,17\}$): The axial, induced pseudoscalar, and pion-nucleon pseudoscalar form factors at all $Q^2$ values for all extraction methods.}
{
\renewcommand{\multirowsetup}{\centering}
\begin{tabular}{|c|c|c|c|c|c|c|}
        \hline
        F7 & \multicolumn{2}{c|}{$G_{\rm A}$} & \multicolumn{2}{c|}{$G_{\rm P}$} & \multicolumn{2}{c|}{$G_{\pi N}$} \\
        \hline
        $Q^{2}$[$\text{GeV}^2$] & Two-state & Summation & Two-state & Summation & Two-state & Summation \\
        \hline
        0.0   & 1.350 (0.067) & 1.353 (0.079) &      -        &      -        & 16.79 (0.838) & 16.83 (0.986) \\
        0.162 & 1.112 (0.082) & 1.066 (0.058) & 23.86 (2.205) & 24.10 (1.577) & 14.70 (1.218) & 12.63 (0.893) \\
        0.315 & 0.991 (0.069) & 0.971 (0.048) & 13.12 (1.015) & 12.78 (0.704) & 13.21 (1.585) & 13.21 (1.020) \\
        0.461 & 0.868 (0.080) & 0.866 (0.058) & 8.250 (0.855) & 7.930 (0.604) & 12.89 (2.908) & 15.16 (1.828) \\
        0.591 & 0.636 (0.093) & 0.645 (0.068) & 4.608 (0.828) & 4.857 (0.606) & 13.06 (2.159) & 11.03 (1.658) \\
        0.725 & 0.708 (0.077) & 0.707 (0.055) & 4.516 (0.516) & 4.535 (0.381) & 11.51 (1.953) & 11.10 (1.216) \\
        0.854 & 0.661 (0.099) & 0.686 (0.070) & 3.676 (0.576) & 3.845 (0.411) & 10.22 (2.981) & 9.864 (1.538) \\
        \hline
\end{tabular}
\label{tab:F7}
}
\end{table*}

\begin{table*}[h!]\small
\caption{G8 ensemble ($a=0.063\,$fm, $m_{\pi}=193\,$MeV, $t_s/a\in\{11,13,15,17\}$): The axial, induced pseudoscalar, and pion-nucleon pseudoscalar form factors at all $Q^2$ values for all extraction methods.}
{
\renewcommand{\multirowsetup}{\centering}
\begin{tabular}{|c|c|c|c|c|c|c|}
        \hline
        G8 & \multicolumn{2}{c|}{$G_{\rm A}$} & \multicolumn{2}{c|}{$G_{\rm P}$} & \multicolumn{2}{c|}{$G_{\pi N}$} \\
        \hline
        $Q^{2}$[$\text{GeV}^2$] & Two-state & Summation & Two-state & Summation & Two-state & Summation \\
        \hline
        0.0   & 1.252 (0.109) & 1.163 (0.086) &      -        &       -       & 14.93 (1.302) & 13.87 (1.020) \\
        0.092 & 1.178 (0.136) & 1.085 (0.062) & 43.19 (6.346) & 43.00 (2.867) & 14.79 (1.464) & 11.11 (0.713) \\
        0.182 & 1.138 (0.119) & 1.059 (0.053) & 26.68 (2.547) & 22.61 (1.167) & 9.900 (2.976) & 15.01 (1.245) \\
        0.268 & 1.060 (0.133) & 1.021 (0.061) & 17.90 (2.238) & 15.87 (1.048) & 7.183 (3.578) & 14.26 (1.346) \\
        0.348 & 0.955 (0.130) & 0.909 (0.059) & 11.49 (1.760) & 11.36 (0.805) & 16.24 (1.973) & 11.72 (0.984) \\
        0.430 & 0.854 (0.109) & 0.876 (0.050) & 9.013 (1.155) & 9.068 (0.534) & 8.564 (1.891) & 11.20 (0.835) \\
        0.509 & 0.776 (0.113) & 0.843 (0.051) & 7.059 (1.048) & 7.449 (0.478) & 6.507 (2.159) & 11.07 (0.816) \\
        \hline
\end{tabular}
\label{tab:G8}
}
\end{table*}

\begin{table*}[h!]\small
\caption{N5 ensemble ($a=0.050\,$fm, $m_{\pi}=429\,$MeV,
$t_s/a\in\{13,16,19,22\}$): The axial, induced pseudoscalar, and pion-nucleon
pseudoscalar form factors at all $Q^2$ values for all extraction methods.}
{
\renewcommand{\multirowsetup}{\centering}
\begin{tabular}{|c|c|c|c|c|c|c|}
        \hline
        N5 & \multicolumn{2}{c|}{$G_{\rm A}$} & \multicolumn{2}{c|}{$G_{\rm P}$} & \multicolumn{2}{c|}{$G_{\pi N}$} \\
        \hline
        $Q^{2}$[$\text{GeV}^2$] & Two-state & Summation & Two-state & Summation & Two-state & Summation \\
        \hline
        0.0   & 1.152 (0.016) & 1.179 (0.027) &       -       &       -       & 16.18 (0.230) & 16.55 (0.372) \\
        0.256 & 0.982 (0.020) & 0.992 (0.022) & 15.24 (0.459) & 15.42 (0.499) & 13.57 (0.331) & 13.66 (0.380) \\
        0.494 & 0.845 (0.018) & 0.874 (0.021) & 8.193 (0.207) & 8.513 (0.243) & 12.63 (0.418) & 12.96 (0.416) \\
        0.719 & 0.745 (0.024) & 0.796 (0.028) & 5.393 (0.197) & 5.608 (0.237) & 11.71 (0.668) & 13.62 (0.693) \\
        0.918 & 0.648 (0.035) & 0.682 (0.041) & 3.910 (0.237) & 4.091 (0.290) & 9.718 (0.696) & 10.53 (0.932) \\
        1.122 & 0.566 (0.031) & 0.614 (0.039) & 2.836 (0.164) & 3.113 (0.214) & 9.377 (0.644) & 9.622 (0.708) \\
        1.317 & 0.532 (0.038) & 0.582 (0.051) & 2.343 (0.169) & 2.538 (0.242) & 8.522 (1.113) & 9.934 (1.040) \\
        \hline
\end{tabular}
\label{tab:N5}
}
\end{table*}

\begin{table*}[h!]\small
\caption{N6 ensemble ($a=0.050\,$fm, $m_{\pi}=331\,$MeV,
$t_s/a\in\{13,16,19,22,25,28\}$): The axial, induced pseudoscalar, and
pion-nucleon pseudoscalar form factors at all $Q^2$ values for all extraction
methods.}
{
\renewcommand{\multirowsetup}{\centering}
\begin{tabular}{|c|c|c|c|c|c|c|}
        \hline
        N6 & \multicolumn{2}{c|}{$G_{\rm A}$} & \multicolumn{2}{c|}{$G_{\rm P}$} & \multicolumn{2}{c|}{$G_{\pi N}$} \\
        \hline
        $Q^{2}$[$\text{GeV}^2$] & Two-state & Summation & Two-state & Summation & Two-state & Summation \\
        \hline
        0.0   & 1.171 (0.025) & 1.206 (0.033) &       -       &      -        & 14.84 (0.317) & 15.29 (0.415) \\
        0.254 & 0.968 (0.030) & 0.981 (0.024) & 14.39 (0.612) & 14.90 (0.486) & 12.76 (0.444) & 12.32 (0.375) \\
        0.487 & 0.785 (0.027) & 0.827 (0.023) & 6.996 (0.288) & 7.459 (0.246) & 11.38 (0.565) & 11.45 (0.452) \\
        0.705 & 0.604 (0.037) & 0.697 (0.033) & 3.877 (0.284) & 4.454 (0.256) & 10.07 (0.961) & 11.83 (0.763) \\
        0.897 & 0.613 (0.054) & 0.700 (0.049) & 3.296 (0.315) & 3.827 (0.288) & 8.780 (1.040) & 8.824 (1.060) \\
        1.092 & 0.472 (0.043) & 0.530 (0.039) & 2.098 (0.202) & 2.367 (0.192) & 7.681 (0.975) & 8.233 (0.818) \\
        1.277 & 0.344 (0.056) & 0.453 (0.053) & 1.333 (0.220) & 1.708 (0.222) & 5.444 (1.721) & 8.951 (1.281) \\
        \hline
\end{tabular}
\label{tab:N6}
}
\end{table*}

\begin{table*}[h!]\small
\caption{O7 ensemble ($a=0.050\,$fm, $m_{\pi}=261\,$MeV, $t_s/a\in\{13,16,19,22\}$): The axial, induced pseudoscalar, and pion-nucleon pseudoscalar form factors at all $Q^2$ values for all extraction methods.}
{
\renewcommand{\multirowsetup}{\centering}
\begin{tabular}{|c|c|c|c|c|c|c|}
        \hline
        O7 & \multicolumn{2}{c|}{$G_{\rm A}$} & \multicolumn{2}{c|}{$G_{\rm P}$} & \multicolumn{2}{c|}{$G_{\pi N}$} \\
        \hline
        $Q^{2}$[$\text{GeV}^2$] & Two-state & Summation & Two-state & Summation & Two-state & Summation \\
        \hline
        0.0   & 1.184 (0.040) & 1.177 (0.044) &       -       &      -        & 13.70 (0.468) & 13.61 (0.504) \\
        0.145 & 0.979 (0.058) & 0.983 (0.035) & 19.94 (1.577) & 20.93 (0.944) & 12.55 (0.665) & 11.57 (0.414) \\
        0.282 & 0.846 (0.051) & 0.896 (0.031) & 10.74 (0.713) & 11.31 (0.448) & 10.94 (0.871) & 11.81 (0.469) \\
        0.412 & 0.743 (0.051) & 0.824 (0.032) & 7.265 (0.531) & 7.486 (0.341) & 7.250 (1.286) & 12.20 (0.670) \\
        0.531 & 0.668 (0.063) & 0.751 (0.038) & 4.981 (0.533) & 5.717 (0.326) & 9.084 (1.066) & 8.901 (0.724) \\
        0.650 & 0.510 (0.052) & 0.654 (0.031) & 3.052 (0.341) & 4.095 (0.208) & 9.321 (0.888) & 8.801 (0.582) \\
        0.764 & 0.455 (0.060) & 0.611 (0.040) & 2.486 (0.330) & 3.231 (0.229) & 5.578 (1.578) & 10.11 (0.730) \\
        \hline
\end{tabular}
\label{tab:O7}
}
\end{table*}

\begin{table}
\caption
{Results for the squared axial radius and the coupling $g_{\rm P}$ on the eleven lattice ensembles,
obtained either with the summation method or the two-state fit.\vspace{16pt}}
{
\label{tab:rAgp}
\begin{tabular}{lrrrr}\hline\hline
 & \multicolumn{2}{c}{Two-state fit} &  \multicolumn{2}{c}{Summation} \\ 
   & $\langle r_{\rm A}^2\rangle$ [fm$^2$] & $g_{\rm P}$ &  $\langle r_{\rm A}^2\rangle$ [fm$^2$] & $g_{\rm P}$ \\
\hline
A3 & 0.137(09) &  1.93(25) & 0.161(16) & 1.72(61) \\
A4 & 0.134(18)  & 2.12(34) & 0.159(29) &  1.66(98) \\
A5 & 0.231(24)  & 3.26(61) & 0.240(28) & 4.82(1.18) \\
B6 & 0.224(47)  & 3.98(63) & 0.262(37) & 5.07(85) 
\\
E5 & 0.129(08)  & 2.03(25) & 0.146(12) & 2.89(56) \\ 
F6 & 0.208(21)  & 2.62(33) & 0.192(20)  & 3.14(58) \\
F7 & 0.233(28)  &  3.76(1.12) & 0.220(23)  & 4.29(80) \\
G8 & 0.279(65)   & 6.31(1.12) & 0.221(39)   &  7.42(1.02)
\\ 
N5 & 0.138(67)  & 1.85(15) & 0.128(09)  & 2.02(31)  \\  
N6 & 0.198(10)   & 2.60(21) & 0.178(10)  & 3.06(41)  \\
O7 & 0.298(21)  &  3.29(30)  & 0.229(15)  & 3.46(42)  \\
\hline\hline
\end{tabular}
}
\end{table}

\clearpage

\bibliographystyle{h-physrev5}
\bibliography{axial_ff_paper}
\end{document}